\documentclass[aps,prb,twocolumn,showpacs,groupedaddress]{revtex4}

\usepackage{amsmath,amssymb}
\usepackage{graphicx}

\begin{document}

\bibstyle{prsty}

\title{Decoherence due to contacts in ballistic nanostructures}

\author{I. Knezevic}
\affiliation{Department of Electrical and Computer Engineering,
University of Wisconsin-Madison, Madison, WI 53706, USA }

\begin{abstract}
The active region of a ballistic nanostructure is an open
quantum-mechanical system, whose nonunitary evolution (decoherence)
towards a nonequilibrium steady state is determined by carrier
injection from the contacts. The purpose of this paper is to provide
a simple theoretical description of the contact-induced decoherence
in ballistic nanostructures, which is established within the
framework of the open systems theory. The active region's evolution
in the presence of contacts is generally non-Markovian. However, if
the contacts' energy relaxation due to electron-electron scattering
is sufficiently fast, then the contacts can be considered memoryless
on timescales coarsened over their energy relaxation time, and the
evolution of the current-limiting active region can be considered
Markovian. Therefore, we first derive a general Markovian map in the
presence of a memoryless environment, by coarse-graining the exact
short-time non-Markovian dynamics of an abstract open system over
the environment memory-loss time, and we give the requirements for
the validity of this map. We then introduce a model contact-active
region interaction that describes carrier injection from the
contacts for a generic two-terminal ballistic nanostructure.
Starting from this model interaction and using the Markovian
dynamics derived by coarse-graining over the effective memory-loss
time of the contacts, we derive the formulas for the nonequilibrium
steady-state distribution functions of the forward and backward
propagating states in the nanostructure's active region. On the
example of a double-barrier tunneling structure, the present
approach yields an I-V curve that shows all the prominent resonant
features. We address the relationship between the present approach
and the Landauer-B\"{u}ttiker formalism, and also briefly discuss
the inclusion of scattering.
\end{abstract}
\date{\today}
\pacs{73.23.-b, 03.65.Yz, 05.60.Gg}

\maketitle


\section{Introduction}\label{sec:Introduction}


In a nanoscale, quasiballistic electronic structure under bias,
relaxation towards a steady state cannot be described by the
semiclassical Boltzmann transport equation \cite{Jacoboni83},
because the structure's active region is typically smaller than the
carrier mean free path and efficient scattering no longer governs
relaxation. Rather, the nanostructure's active region behaves as an
open quantum-mechanical system, \cite{Potz89,Frensley90} exchanging
particles with the reservoirs of charge  (usually referred to as
leads or contacts). In the absence of scattering within the active
region, the coupling of the active region to the contacts is the
cause of its nonunitary evolution (decoherence) towards a
nonequilibrium steady state, and the importance of
this coupling has become well-recognized in quantum transport
studies. The description and manipulation of the contact-induced
decoherence are presently of great importance not only in quantum
transport studies,
\cite{Bird03,Elhassan05,Grubin98,Ferry03,Svizhenko03,Knezevic03_3,Ferrari04,Gebauer04,Bushong05}
but also in the theory of measurement \cite{Zurek03} and quantum
information. \cite{Wu04}

The purpose of this paper is to provide a simple description of the
nonunitary evolution of a ballistic nanostructure's active region
due to the injection of carriers from the contacts. Carrier injection from the contacts into the active region is
traditionally described by either an explicit source term, such as
in the single-particle density matrix,
\cite{Jacoboni92,Rossi92,Brunetti89,Hohenester97,Ciancio04,Platero04}
Wigner function
\cite{Kluksdahl89,Potz89,Frensley90,Bordone99,Biegel97,Jensen89,Grubin02,Shifren03,Jacoboni03,Nedjalkov04,Nedjalkov06}
and Pauli equation \cite{Fischetti98,Fischetti99} transport
formalisms, or via a special self-energy term in the ubiquitous
nonequilibrium Green's function formalism.
\cite{Lake92_1,Lake97,Datta92_1,Datta92_2,Mamaluy05,Svizhenko02} In
this work, the problem of contact-induced decoherence is treated
using the open systems formalism: \cite{Breuer02} we start with a
model interaction Hamiltonian that describes the injection of
carriers from the contacts, and then deduce the resulting nonunitary
evolution of the active region's many-body reduced statistical
operator in the Markovian approximation. The following two features
distinguish this paper from other recent works,
\cite{Gurvitz96,Gurvitz97,Li05,Pedersen05} in which Markovian rate
equations have also been derived for tunneling nanostructures:

(1) Derivation of the Markovian evolution is achieved by coarse
graining of the exact short-time dynamics in the presence of
memoryless contacts, rather than utilizing the weak-coupling and van
Hove limit, \cite{Li05,Pedersen05} or the high-bias limit.
\cite{Gurvitz96} Namely, electron-electron interaction is typically
the leading inelastic scattering mechanism in the contacts. If the
contacts' energy-relaxation time $\tau$ due to electron-electron
scattering is sufficiently short, then on the timescales coarsened
over $\tau$, the contacts appear memoryless and the evolution of the
current-limiting active region can be considered Markovian.
\cite{Rossi02,Wingreen90} The approximation of a memoryless
environment, as applied to nanostructures, will be discussed in
detail in Sections \ref{sec:Decoherence in the presence of a
"memoryless" environment} and \ref{sec:Model}.

(2) A model contact-active region interaction is introduced to
describe the injection of carriers through the open boundaries and
supplant the resonant level model. Namely, for tunneling
nanostructures, like a resonant-tunneling diode, it is common to
adopt the resonant-level model \cite{Jauho84} when trying to
separate the active region from the contacts: the active region is
treated as a system with one or several discrete resonances. But the
resonant-level model for the active region is inapplicable away from
the resonances, and cannot, for instance, capture the current
increase in a resonant-tunneling diode at high biases (larger than
the valley bias) that is due to the continuum states. Also, it is
not a good model for simple structures without resonances, such as
an $nin$ diode or the channel of a MOSFET. So we introduce an
alternative model Hamiltonian that does not assume resonances
\textit{a priori} exist and that works both near and far from
resonances. It captures the open boundaries and naturally continuous
spectrum of a nanostructure's active region, and describes carrier
injection in a manner conceptually similar to the explicit source
terms in the single-particle density matrix or Wigner function
techniques.

The paper is organized as follows: in Sec. \ref{sec:Formalism}, we
overview the basics of the partial-trace-free formalism
\cite{Knezevic02} for the treatment of open systems
(\ref{sec:Decomposition}) and present the main steps in the
derivation of the non-Markovian equations with memory dressing
(\ref{sec:Memory Dressing}). \cite{Knezevic04} In Sec.
\ref{sec:Decoherence in the presence of a "memoryless" environment},
we discuss how the fast memory loss due to electron-electron
scattering in the contacts can be used to justify a Markovian
approximation to the exact evolution of the active region in a small
semiconductor device or a ballistic nanostructure. In Sec.
\ref{sec:Coarse Graining}, we then perform coarse-graining of the
exact non-Markovian short-time dynamics of an abstract open system
(details of the derivation of the exact short-time dynamics are
given in Appendix A) over the memory loss time of the environment in
order to obtain a Markovian map, and we discuss the necessary
conditions for this procedure to hold. In Sec. \ref{sec:Model}, we
introduce a model contact-active region interaction applicable to a
generic two-terminal nanostructure, which describes carrier
injection from the contacts. This model interaction does not require
that the structure \textit{a priori} possesses resonances. In Sec.
\ref{sec:CurrentCarryingContactsMemoryless}, we formalize the
requirements for the current-carrying contacts to be considered a
memoryless environment. Starting from the model interaction and
using the Markovian dynamics derived, we then proceed to derive the
Markovian evolution and steady-state values for the distribution
functions of the forward and backward propagating states in the
active region of a nanostructure, and we also give the result for
the steady-state current (Sec. \ref{sec:Lambda for model and steady
state distributions}). We discuss the relationship of the presented
approach to the Landauer-B\"{u}ttiker formalism
\cite{Landauer57,Landauer70,Buttiker85,Buttiker86} in Sec.
\ref{sec:LB comparison}. In Sec. \ref{sec:RTD}, we work out the
example of a one-dimensional double-barrier tunneling structure. The
nonequilibrium steady states obtained as a result of the Markovian
evolution at different biases produce an I-V curve that shows all
the prominent resonant features, and we compare the results to those
predicted by the Landauer-B\"{u}ttiker formalism. The manuscript is
concluded in Sec. \ref{sec:Conclusion}, with a brief summary and
some final remarks on the inclusion of scattering and lifting the
Markovian approximation.

\section{The Formalism}\label{sec:Formalism}

\subsection{Decomposition of the Liouville space}\label{sec:Decomposition}

Let us consider an open system $S$, coupled with the environment
$E$, so that the composite $SE$ is closed. For a ballistic nanostructure, $S$ would represent the active region, while $E$ would be the contacts; more generally,
if scattering due to phonons occurs within the active region, phonons should also be included as part of $E$. \cite{Fischetti99}
$S$, $E$, and $SE$ are assumed
to have finite-dimensional Hilbert spaces, of dimensions $d_S$,
$d_E$, and $d_Sd_E$, respectively. Consequently, their Liouville
spaces -- the spaces of operators acting on the above Hilbert spaces
-- are of dimensions $d_S^2$, $d_E^2$, and $d_S^2d_E^2$,
respectively. The total $SE$ Hamiltonian $\mathcal H$ is generally a
sum of a system part $1_E\otimes \mathcal H_S$, an environment part
$ {{\mathcal{H}}}_E\otimes 1_S$, and an interaction part ${\mathcal
H}_{\mathrm{int}}$. The total Hamiltonian $\mathcal H$ (acting on
the $SE$ Hilbert space) induces the total $SE$ Liouvillian $\mathcal
L$ (acting on the $SE$ Liouville space) through the commutator,
which governs the evolution of the $SE$ statistical operator $\rho$
according to the Liouville equation
\begin{equation}\label{Liouville}
\frac{d\rho}{dt}=-i\left[\mathcal H ,\rho\right]=-i\mathcal L\rho .
\end{equation}
$\mathcal H$ and $\mathcal L$ are given in the units of frequency.
Dynamics of the open system $S$ is described by its reduced statistical operator $\rho_S$, obtained from $\rho$ by tracing out the environment
states
\begin{equation}
\rho_S=\mathrm{Tr}_E\rho .
\end{equation}
In general, the dynamics of $\rho_S$ is not unitary. A common
approach to calculating the evolution of $\rho_S$ is by using
projection operators \cite{Breuer02,Nakajima58,Zwanzig60} that act
on the $SE$ Liouville space. Typically, an environmental statistical operator $\rho_E$ is chosen to induce a projection operator $P$ by
$P\mu=\rho_E\otimes\mathrm{Tr}_E \mu$, where $\mu$ is any vector
from the $SE$ Liouville space. Coupled equations of motion for
$P\rho$ and $(1-P)\rho$ are then solved, often in the weak-coupling
limit, and the reduced dynamics is obtained from
$\rho_S=\mathrm{Tr}_E\rho=\mathrm{Tr}_E(P\rho)$.

Most often, the projection operator utilized is induced by the
initial environmental statistical operator $\rho_E(0)$ \cite{Breuer02}.
The reason is that, in the most common approximation of initially
decoupled $S$ and $E$, described by
$\rho(0)=\rho_E(0)\otimes\rho_S(0)$, the projection operator induced
by $\rho_E(0)$ will eliminate a certain memory term occurring in the
evolution of $\rho_S$. However, the result for the final dynamics
must not depend on the projection operator used, as projection
operators are, after all, only auxiliary quantities. In this paper,
we will follow the work on the partial-trace-free approach of  Ref.
\onlinecite{Knezevic02}, that uses the projection operator
$\overline P$ induced by the uniform environment statistical operator
$\overline\rho_E=d_E^{-1}\mathrm{diag}(1\dots 1)$. $\overline P$ has
a unique property: it is the only projection operator that has an
orthonormal eigenbasis in which it is represented by a diagonal
form. Its unit eigenspace, of dimension $d_S^2$, is a mirror-image
of the Liouville space of the open system $S$. \textit{Projecting
onto the unit eigenspace of $\overline P$ is equivalent to taking
the partial trace with respect to environmental states}
\cite{Knezevic02}, because for any element of the $SE$ Liouville
space it holds
\begin{equation}\label{isomorphism}
\left(\overline P\mu\right)^{\overline{\alpha\beta}}=
d_E^{-1/2}\left(\mathrm{Tr}_E \mu\right)^\mathrm{{\alpha\beta}}.
\end{equation}
Here, the unit-eigenspace of $\overline P$ is spanned by a basis
$|\overline{\alpha\beta}\rangle$, while the Liouville space of $S$
is spanned by $|{\alpha\beta}\rangle$, where the two bases are
isomorphic through the following simple relationship
\begin{equation}\label{overline P spectral form}
\left | \overline{\alpha\beta} \right\rangle
=d_E^{-1/2}\sum_{j=1}^{d_E} \left | {j\alpha ,j\beta} \right\rangle.
\end{equation}
$\left | {i\alpha ,j\beta} \right\rangle$ is a basis in the $SE$
Liouville space, induced by the bases $\left |i\, j \right\rangle$
and $\left | \alpha\beta \right\rangle$ in the environment and
system Liouville spaces, respectively.

\begin{figure}
\center\includegraphics[width=5cm]{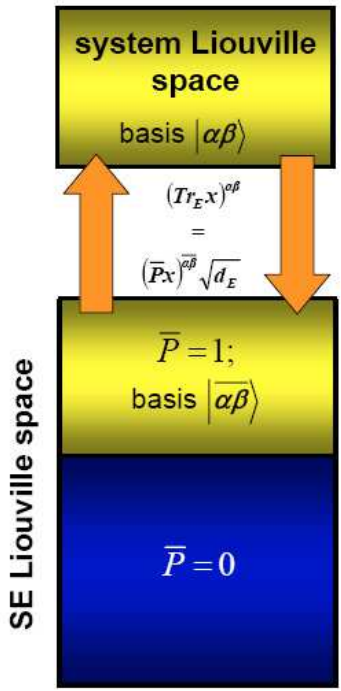}
\caption{\label{fig:Decomposition} (Color online) Decomposition of the total $SE$
Liouville space into the eigenspaces of the projection operator
$\overline P$, induced by the uniform statistical operator $\overline
\rho_E$. The unit eigenspace is equivalent to the system Liouville
space, where the equivalence is given by the isomorphism
(\ref{isomorphism}). Working within the eigenspaces of $\overline P$
removes the need for performing the partial trace over the
environmental states.}
\end{figure}

Decomposition of the $SE$ Liouville space into the two eigenspaces
of $\overline P$ (depicted in Fig. \ref{fig:Decomposition}) is the
essence of the PTF approach: every vector $\mu$ from the $SE$
Liouville space can be written as a column $\mu=[\mu_1\;
\mu_2]^\mathrm{T}$, where $\mu_1$ belongs to the unit eigenspace of
$\overline P$ and represents (up to a multiplicative constant
$\sqrt{d_E}$) the system's reduced component of $\mu$, i.e.,
$\mu_S\equiv \mathrm{Tr}_E \mu=\sqrt{d_E}\mu_1$. The other
component, $\mu_2$, belongs to subspace 2 (the zero-eigenspace of
$\overline P$), where the correlations between $S$ and $E$ reside.
It is important to note that the elements of subspace 2 (blue
subspace in Fig. \ref{fig:Decomposition}) have zero trace over
environmental states.

In a similar fashion, an operator $A$ acting on the $SE$ Liouville
space has a block-form with submatrices $A_{pq}$, $p,q=1,2$, where
$A_{11}$ would be the system's reduced component of this operator.
For instance, the block form of the $SE$ Liouvillian $L$ is given by
\begin{equation}
\mathcal{L}=\left[\begin{array}{cc}
{\mathcal{L}}_{11} & {\mathcal{L}}_{12} \\
{\mathcal{L}}_{21}& {\mathcal{L}}_{22} \end{array} \right],
\end{equation}
where  ${\mathcal{L}}_{11}$ is commutator-generated, and corresponds
to an effective system Hamiltonian ${\mathcal{H}}_S+\mathrm{Tr}_E (
{{\mathcal{H}}}_{\mathrm{int}})/d_E$. Off-diagonal, non-square
Liouvillian submatrices, ${\mathcal{L}}_{12}$ and
${\mathcal{L}}_{21}={\mathcal{L}}_{12}^\dagger$, represent the
$S$--$E$ interaction as seen in the composite Liouville space --
when ${\mathcal{H}}_{\mathrm{int}}$ vanishes, so do
${\mathcal{L}}_{12}$ and ${\mathcal{L}}_{21}$. ${\mathcal{L}}_{22}$
can be perceived as governing the evolution of entangled $SE$
states, and tends to a form fixed by $ {{\mathcal{H}}}_S$ and
${\mathcal{H}}_E$ when the interaction is turned off.

\subsection{Equations with
memory dressing}\label{sec:Memory Dressing}
Using the notation introduced above, the evolution of the reduced
statistical operator $\rho_S$ can be represented by
\begin{equation}\label{rhoSbasic}
\rho_S(t)=\mathcal{U}_{11}(t,0)\rho_S(0)+\sqrt{d_E}\,
\mathcal{U}_{12}(t,0)\rho_2(0),
\end{equation}
where $\mathcal{U}_{11}$ and $\mathcal{U}_{12}$ are the submatrices
od the $SE$ evolution operator $\mathcal U$, given by
\begin{eqnarray}
\mathcal U(t,0)&=&\mathrm{T^c}\exp\left(-i\int_0^t
\left[\begin{array}{cc}
{\mathcal{L}}_{11} & {\mathcal{L}}_{12} \\
{\mathcal{L}}_{21}& {\mathcal{L}}_{22} \end{array} \right]\,
dt\right)\nonumber\\
&=&\left[\begin{array}{cc}
\mathcal U_{11}(t,0) & \mathcal U_{12}(t,0) \\
\mathcal U_{21}(t,0)& \mathcal U_{22}(t,0)\end{array}
\right].
\end{eqnarray}
In Ref. \onlinecite{Knezevic04}, equations of motion for
$\mathcal{U}_{11}$ and $\mathcal{U}_{12}$ were derived as
\begin{subequations}\label{Us}
\begin{eqnarray}
\frac{d{\mathcal
U}_{11}}{dt}&=&-i\left({\mathcal{L}}_{11}-{\mathcal{L}}_{12}\mathcal{R}\right)\mathcal
U_{11},\hfill\\
\frac{d{\mathcal
U}_{12}}{dt}&=&-i\left({\mathcal{L}}_{11}-{\mathcal{L}}_{12}\mathcal{R}\right)\mathcal
U_{12}-i{\mathcal{L}}_{12}\mathcal V,\hfill
\end{eqnarray}
\end{subequations}

\noindent accompanied by the initial conditions
$\mathcal{U}_{11}(0,0)=1$ and $\mathcal{U}_{12}(0,0)=0$. Quantity
$\mathcal{R}$ is the so-called \textit{memory dressing}, as it
appears to "dress" the real physical interaction
${\mathcal{L}}_{12}$ and yield an effective (generally complex)
interaction term, $-{\mathcal{L}}_{12}\mathcal{R}$, that accompanies
the hermitian term ${\mathcal{L}}_{11}$, responsible for unitary
evolution. Memory dressing describes the cumulative effect of the
$S-E$ interaction, as witnessed by a quadratic feedback term in its
self-contained matrix Riccati \cite{Reid72,Bittani91} equation  of
motion (below). The other new quantity occurring in (\ref{Us}),
$\mathcal V(t,0)$, can be perceived as the evolution operator for
the states from subspace 2, and is important for the description of
the influx of information from $E$ to $S$. $\mathcal{R}$ and
$\mathcal V$ obey

\begin{subequations}\label{RandV}
\begin{eqnarray}
\frac{d\mathcal{R}}{dt}&=&-i {\mathcal{L}}_{22}\mathcal{R}-i\mathcal{R}{\mathcal{L}}_{12}\mathcal{R}+i\mathcal{R}{\mathcal{L}}_{11}+i {\mathcal{L}}_{21},\hfill\label{R}\\
\frac{d\mathcal{V}}{dt}&=&-i \left(
{\mathcal{L}}_{22}+\mathcal{R}{\mathcal{L}}_{12}\right)\mathcal{V},\label{V}
\end{eqnarray}
\end{subequations}

\noindent accompanied by $\mathcal{R}(0)=0$ and
$\mathcal{V}(0,0)=1$.

Equations (\ref{Us}) and (\ref{RandV}) are
\textit{exact}:\cite{Knezevic04} they are an alternative form of the
$SE$ Liouville equation (\ref{Liouville}). The resulting exact
evolution of the reduced statistical operator can be expressed through the
following differential equation of motion
\begin{eqnarray}\label{dot rho_S}
\frac{d\rho_S(t)}{dt}=&-&i\left[{\mathcal{L}}_{11}-{\mathcal{L}}_{12}\mathcal{R}(t)\right]\rho_S(t)\nonumber\\
&-&i{\mathcal{L}}_{12}\sqrt{d_E}\mathcal V(t,0)\mathcal \rho_2(0).
\end{eqnarray}
which is a partial-trace-free form of
$\frac{d\rho_S}{dt}=\mathrm{Tr}_E (-iL\rho)$.

If we restrict our attention to the evolution starting from an
initially uncorrelated state of the form
\begin{equation}\label{uncorrelated rho_s}
\rho(0)=\rho_E(0)\otimes\rho_S(0),
\end{equation}
it is possible to completely reduce the problem to subspace $1$.
Namely, it is possible to write
\begin{equation}\label{rho2vsrhoS}
\rho_2(0)=\mathcal{M}\rho_1(0)=d_E^{-1/2}\mathcal{M}\rho_S(0),
\end{equation}
where the mapping $\mathcal{M}$ is completely determined by the
components of $\rho_E(0)$, the initial environment statistical operator
(see Appendix \ref{appendix:uncorr}). Equation (\ref{rho2vsrhoS})
embodies the argument made by Lindblad \cite{Lindblad96} that a
subdynamics exists only for an uncorrelated initial state, because,
as a consequence of (\ref{rhoSbasic}) and (\ref{rho2vsrhoS}), it is
possible to write
\begin{eqnarray}\label{eq:integral uncorrelated}
\rho_S(t)=\underbrace{[\mathcal{U}_{11}(t,0)+\mathcal{U}_{12}(t,0)
\mathcal{M}]}_{\mathcal{W}(t,0)}\rho_S(0),
\end{eqnarray}
so the evolution is completely described on the Liouville space of
the open system. When (\ref{rho2vsrhoS}) is
substituted into (\ref{dot rho_S}), we obtain the differential form of (\ref{eq:integral uncorrelated}) as
\begin{equation}\label{eq:dot rho_S uncorrelated}
\frac{d\rho_S(t)}{dt}=-i\left[{\mathcal{L}}_{11}-{\mathcal{L}}_{12}\mathcal{R}(t)\right]\rho_S(t)-i{\mathcal{L}}_{12}\mathcal
V(t)\mathcal M\rho_S(0).
\end{equation}

\noindent It is well known that a subdynamics can also be
obtained for the case of an initially decoupled $SE$ state by simply
choosing the initial environmental statistical operator $\rho_E$ as the
one to induce the projection operator $P$ (see, for instance, Ref.
\onlinecite{Breuer02}). However, the result for the final dynamics must
not depend on the projection operator used, as projection operators
are, after all, only auxiliary quantities. While the physics must be the same regardless of the projection
operator used, the opacity of the equations
obtained certainly varies. Equation (\ref{eq:dot rho_S uncorrelated})
shows explicitly \textit{how} the subdynamics looks for $\overline
P$; by generalizing the proof in Appendix \ref{appendix:uncorr}, one
can write the subdynamics for any other projection operator instead
of $\overline P$. The reason we are using $\overline P$ instead of
the projection operator $P$ induced by the initial environmental
statistical operator is that, as stated previously, $\overline P$ is the
only projection operator that has an orthonormal eigenbasis in which
it is represented by a diagonal form (\ref{overline P spectral
form}). While any other projection operator $P$ still projects onto
its own $d_S^2$-dimensional image space (see Appendix
\ref{appendix:uncorr}), $P$ and $1-P$ never assume simple diagonal
forms, so after projecting one still needs to explicitly take the
partial trace, which leaves the equations less transparent.

\section{Decoherence in the presence of a "memoryless" environment}
\label{sec:Decoherence in the presence of a "memoryless" environment}

The non-Markovian map $\mathcal{W}(t,0)=\mathcal{U}_{11}(t,0)+\mathcal{U}_{12}(t,0)\mathcal{M}$ that defines
the subdynamics (\ref{eq:integral uncorrelated}) can quite generally be written as
\begin{equation}\label{eq:W in terms of K}
\mathcal{W}(t,0)=\mathrm{T^c}\exp{\left(\int_0^t \mathcal{K}(t')\,dt'\right)}.
\end{equation}
Here, $\mathcal{K}(t)$ is the generator of $\mathcal{W}(t,0)$. In
general,
$\mathcal{K}(t)=-i\mathcal{L}_{\mathrm{eff}}-\mathcal{G}(t)$, i.e.,
it contains an effective system Liouvillian
$\mathcal{L}_{\mathrm{eff}}$ and a correction $\mathcal G$ due to
the system-environment interaction, which describes decoherence. In
case of Markovian evolution,
$\mathcal{K}=-i\mathcal{L}_{\mathrm{eff}}-\mathcal{G}=\mathrm{const.}$,
and $\mathcal{G}$ must have the well-known Lindblad dissipator form
\cite{Lindblad76,Alicki87} in order for the map (\ref{eq:W in terms
of K}) to remain completely positive. \cite{Alicki87,Breuer02}

In general, it is impossible to obtain $\mathcal{W}(t,0)$ exactly. If one is interested in retaining the non-Markovian nature of
(\ref{eq:W in terms of K}), typically an expansion up to the second or fourth order in the
interaction is undertaken. \cite{Breuer02} On the other hand, a Markovian approximation to the exact dynamics
can be obtained in the weak-coupling and van Hove limit, as first shown by
Davies. \cite{Davies74}
Although the weak coupling limit has been used previously by several authors \cite{Li05,Pedersen05}
to derive Markovian rate equations for tunneling structures in the resonant-level model,
this approximation is not generally applicable for nanostructures. \cite{Li05}

The point we wish to make here is that the Markovian approximation
to the long-time evolution of nanostructures can be justified more
broadly, by employing \textit{the approximation of a memoryless
environment} for the contacts. Consider first the active region of a
small semiconductor device; a good example is the state-of-the art
MOSFET with 45 nm lithographic gate length (physical gate length is
estimated to be around 20 nm), found in Intel's 2008 Penryn
processors. \cite{IntelPenryn} Semiconductor devices are generally
required to operate at (or at least near) room temperature, where
phonons are abundant. However, due to the active region's minuscule
dimensions, scattering happens infrequently, so the active region
does feature quasiballistic transport, where scattering can be added
as a perturbation to the ballistic solution. The bulk-like contacts
of semiconductor devices are typically heavily doped (e.g., $\sim
10^{19}- 10^{20}\;\mathrm{cm}^{-3}$ in silicon), and at room
temperature all the dopants are ionized; at such high doping
densities, electron-electron scattering dominates over phonon
scattering as the leading energy-relaxation mechanism (e.g.,
relaxation time for electron-electron scattering in bulk GaAs at
$10^{19}\;\mathrm{cm}^{-3}$ and room temperature is $10$ fs,
\cite{Kriman90} whereas it is about $150$ fs for polar optical
phonon scattering \cite{Lundstrom00}). Basically, electron-electron
scattering in the highly doped contacts of semiconductor devices
ensures that the carrier distribution snaps into a distribution that
can be considered a displaced (also known as drifted) Fermi-Dirac
distribution \cite{Lugli85} (see also Sec.
\ref{sec:CurrentCarryingContactsMemoryless}) within the
energy-relaxation time $\tau\approx 10^1-10^2$ femtoseconds
\cite{Osman87,Kriman90} (the actual value depends on the doping
density and temperature). This time is very short with respect to
the typical response times of these devices, which is on the
timescales of $\tau_{AR}\approx 1-10$ ps ("AR" stands for the active
region). Therefore, for small semiconductor devices, on timescales
coarsened over the energy relaxation time $\tau$ of the contacts,
the contact distribution function responds virtually
instantaneously, and the contacts can be considered
\textit{memoryless}, while the relaxation of the whole structure
happens on timescales a few orders of magnitude longer. (A
memoryless approximation must be applied with care to
current-carrying contacts, as we will discuss in detail in
\ref{sec:CurrentCarryingContactsMemoryless} and \ref{sec:LB
comparison}.)

For low-dimensional nanostructures, fabricated on a high-mobility
two-dimensional electron gas (2DEG) and operating at low
temperatures, the energy relaxation in the contacts is also governed
by the inelastic electron-electron scattering,
\cite{Altshuler79,Altshuler81} because the phonons are frozen
(although there are indications that acoustic phonon scattering may
be important down to about 4 K \cite{Lutz93}). The near-equilibrium
energy relaxation times in these contacts are much longer than in
devices, falling in the wide range of $10^0-10^3$ ps,
\cite{Giuliani82,Fasol91,Noguchi96,Kabir06} depending on the contact
dimensionality (1D\cite{Altshuler82,Wind86,Linke97,LeHur06} or 2D
\cite{Giuliani82,Fasol91}), carrier density, and temperature.
Excitations with energies higher than $k_BT$, such as when bias
$V>k_BT/q$ is applied across the nanostructure ($q$ is the electron
charge), relax more rapidly, \cite{Fasol91,Linke97} which is of
particular importance in the collector contact. In low-dimensional
nanostructures, there are also experimental indications that
coupling of the active region to the contacts governs its evolution.
\cite{Pivin99} As for the typical response times of nanostructures,
recent experiment by Naser \textit{et al.} \cite{Naser06}
demonstrated Markovian relaxation in quantum point contacts on
$\tau_{AR}\approx 50$ ns timescales at 4 K, so the ratio
$\tau/\tau_{AR}$ is still less than unity, but not as small as in
devices.

Still, there is enough rationale to further explore a
nanostructure's dynamics within the approximation of memoryless
contacts, with the understanding that this approximation must
generally be qualified, especially for nanostructures at very low
temperatures. We will therefore proceed with deriving the Markovian
approximation to the exact non-Markovian equation (\ref{eq:dot rho_S
uncorrelated}) in the presence of an environment that loses memory
on a timescale $\tau$, presumed much shorter that the response time
of the open system, and we will derive the relationships that the
coarse graining time $\tau$ must satisfy for the approximation to be
consistent. Then, in Sec. \ref{sec:Model}, we will see what type of
constraint that puts on our energy relaxation time in the contacts.

Before proceeding with the formal development, it is worth stressing
that the importance of a Markovian approximation to the exact
evolution is great, because with both nanoscale semiconductor
devices used for digital applications and with DC experiments on
nanostructures, \textit{one is primarily interested in the steady
state} that the structure reaches upon the application of a DC bias.
In these situations, it is sufficient to employ the Markovian
approximation to the evolution (if warranted), as it is correct on
long timescales and will result in the correct steady state.

\subsection{Markovian evolution by coarse
graining}\label{sec:Coarse Graining}

To practically obtain the Markovian approximation due to an
environment that loses memory after a time $\tau$, we use the coarse-graining procedure:
we can partition the time
axis into intervals of length $\tau$, $t_n=n\tau$, so the
environment interacts with the system in exactly the same way during
each interval $[t_n,t_{n+1}]$, \cite{Lidar01} so
\begin{equation}\label{eq:difference rho_S}
\frac{\rho_{S,n+1}-\rho_{S,n}}{\tau}=
\mathcal{\overline{K}}_\tau\rho_{S,n},\end{equation}
where $\mathcal{\overline{K}}_\tau=\frac{\int_0^\tau \mathcal{K}(t')dt'}{\tau}=\frac{\int_{t_n}^{t_{n+1}}\mathcal{K}(t')dt'}{\tau}$
is the averaged value of the map's generator over any interval $[t_n,t_{n+1}]$ ($\mathcal{K}$ is reset at each $t_n$).
If the timescales are coarsened over $\tau$, then the term on the left of (\ref{eq:difference rho_S})
approximates the first derivative at $t_n$, so
the system's evolution can be described by
\begin{equation}\label{eq:Markov_final0}
\frac{d\rho_S}{dt}=\mathcal{\overline{K}}_\tau\rho_S(t).
\end{equation}
The above map is completely positive and Markovian (coarse graining preserves complete positivity \cite{Lidar01}),
but still has little practical value, because extracting $\mathcal{K}$ explicitly from first principles is
difficult. However, if the coarse-graining time $\tau$ is short enough, then the short-time expansion of $\mathcal{K}$ can be used to perform the coarse-graining.
Up to the second order in time (details of the short-time expansion can be found in Appendix \ref{sec:STE map}),
\begin{equation}\label{eq:short time expansion of K}
\mathcal{K}(t)=-i\mathcal{L}_{\mathrm{eff}}-2\Lambda t +o(t^2),
\end{equation}
where
${\mathcal{L}}_{\mathrm{eff}}=[{\mathcal{H}}_S+\langle{{\mathcal{H}}_\mathrm{{int}}}\rangle,\dots]
=\mathcal L_S+[\langle{{\mathcal{H}}_\mathrm{{int}}}\rangle,\dots]$
is an effective system Liouvillian, containing the
noninteracting-system Liouvillian $\mathcal L_S$ and a correction
due to the interaction [$\langle\dots\rangle=\mathrm{Tr}_E[\rho_E(0)\dots]$ denotes the
partial average with respect to the initial environmental state
$\rho_E(0)$]. The matrix elements of superoperator $\Lambda$, in a
basis $\alpha\beta$ in the system's Liouville space (Liouville space
is basically a tensor square of the Hilbert space), are determined
from the matrix elements of the interaction Hamiltonian:
\begin{eqnarray}\label{eq:Lambda}
&{\Lambda}^{\alpha\beta}_{\alpha '\beta'}=\frac{1}{2}\left\{
\left\langle{{\mathcal{H}}}_{\mathrm{int}}^2\right\rangle^{\alpha}_{\alpha
'}\delta^{\beta '}_\beta +\left\langle
{{\mathcal{H}}}_{\mathrm{int}}^2\right\rangle^{\beta'}_{\beta
'}\delta^\alpha_{\alpha '}\right.\quad\hfill\nonumber\\
&-2\sum_{j,j'}\left(
{{\mathcal{H}}}_{\mathrm{int}}\right)^{j'\alpha}_{j\alpha
'}\rho_E^j\left(
{{\mathcal{H}}}_{\mathrm{int}}\right)^{j\beta'}_{j'\beta}\quad\quad\hfill\\
&-\left.\left(\langle
{\mathcal{H}}_{\mathrm{int}}\rangle^2\right)^{\alpha}_{\alpha
'}\delta^{\beta'}_\beta +2\langle
{\mathcal{H}}_{\mathrm{int}}\rangle^{\alpha}_{\alpha'}\langle
{\mathcal{H}}_{\mathrm{int}}\rangle^{\beta'}_\beta -\left(\langle
{\mathcal{H}}_{\mathrm{int}}\rangle^2\right)^{\beta'}_\beta\delta^{\alpha}_{\alpha
'}\right\},\nonumber
\end{eqnarray}
where $\rho_E^j$ are the eigenvalues of the initial environment
statistical operator $\rho_E(0)$. $\Lambda$ has implicitly been defined
previously \cite{Alicki89} in the interaction picture and with the
assumption of $\langle {\mathcal{H}}_{\mathrm{int}}\rangle =0$.
Here, we work in the Schr\"{o}dinger picture and generally need to
retain $\langle {\mathcal{H}}_{\mathrm{int}}\rangle\neq 0$, which is
important for the inclusion of carrier-carrier interaction in
nanostructures. ${\Lambda}$ contains essential information on the
directions of coherence loss.

If the coarse-graining time $\tau$ is short enough that it holds
\begin{equation}\label{eq:validity of Markov1}
\left\|\Lambda\right\|\tau \ll \left\|\mathcal{L}_{\mathrm{eff}}\right\|,
\end{equation}
then the short-time expansion of $\mathcal K$ can be used for coarse-graining, and we obtain
\begin{equation}\label{eq:coarsened K}
\mathcal{\overline{K}}_\tau =-i\mathcal{L}_{\mathrm{eff}}-\Lambda\tau,
\end{equation}
leading to the Markovian equation
\begin{equation}\label{eq:Markov final}
\frac{d\rho_S(t)}{dt}=\left(-i\mathcal{L}_{\mathrm{eff}}-\Lambda\tau\right)\rho_S(t),
\end{equation}
which is the central equation of this paper. For the Markovian
approximation to be consistent,  \cite{Breuer02} the system's relaxation (occurring
on timescales no shorter than $1/\|\Lambda\|\tau$) must be much slower than the
environment's relaxation (occurring over $\tau$), therefore we must have
\begin{equation}\label{eq:validity of Markov2}
\left\|\Lambda\right\| {\tau}^2 \ll 1.
\end{equation}

Conditions (\ref{eq:validity of Markov1}) and (\ref{eq:validity of
Markov2}) can compactly be written as
\begin{equation}\label{eq:validity of Markov}
\left\|\Lambda\right\| {\tau}^2 \ll
\min{\{1,\left\|\mathcal{L}_{\mathrm{eff}}\right\|\tau\}}.
\end{equation}

\subsection{Some general considerations regarding the use of Eq. (\ref{eq:Markov final})}\label{sec:Misc Considerations}
Before we proceed to treating a concrete nanostructure as an example,
there are several general features regarding the use of Eq. (\ref{eq:Markov final}) that can be used more broadly than in the
treatment of nanostructures. (The reader interested exclusively in decoherence in nanostructures can skip the rest of this section and go directly to Sec. \ref{sec:Model}.)

\subsubsection{Decoherence-free evolution in the zero-eigenspace of
${\Lambda}$ }

Let us assume for a moment that $\mathcal{L}_{\mathrm{eff}}$ and
${\Lambda}$ commute (we will see two cases of this situation in Appendix
\ref{sec:examples}). If so, the components of $\rho_S$
belonging to the null-space of ${\Lambda}$ will not decohere -- they
will continue to evolve unitarily, as the null-space of ${\Lambda}$
will be invariant under $\mathcal{L}_{\mathrm{eff}}$. Components of
$\rho_S$ corresponding to the nonzero eigenvalues of ${\Lambda}$
will decohere until they drop to zero. So in the case of commuting
$\mathcal{L}_{\mathrm{eff}}$ and ${\Lambda}$, null-space of
${\Lambda}$ is decoherence-free. For non-commuting
$\mathcal{L}_{\mathrm{eff}}$ and ${\Lambda}$, this statement can be
generalized to

\textbf{Theorem 1.} \textit{If a subspace of $\mathcal
N({\Lambda})$, the null-space of operator ${\Lambda}$, is also an
invariant subspace of $\mathcal{L}_{\mathrm{eff}}$, then it supports
decoherence-free (unitary) evolution according to the map
(\ref{eq:Markov final}).}

\textit{Proof}. Let $\mathcal N '({\Lambda})$ be a subspace of
$\mathcal N({\Lambda})$. If $\mathcal N '({\Lambda})$ is an
invariant subspace of $\mathcal{L}_{\mathrm{eff}}$, then it is an
invariant subspace of the full generator of the Markovian semigroup
(\ref{eq:Markov final}), and consequently an invariant subspace of the
semigroup. A statistical operator $\rho^0$, initially prepared in
$\mathcal N '({\Lambda})$, would remain in $\mathcal N '({\Lambda})$
at all times, and evolve unitarily according to
$\frac{d\rho^0}{dt}=-i\mathcal{L}_{\mathrm{eff}}^0\rho^0(t)$, where
$\mathcal{L}_{\mathrm{eff}}^0$ is the reduced form of
$\mathcal{L}_{\mathrm{eff}}$ onto $\mathcal N'({\Lambda})$.
$\blacksquare$

This theorem is equivalent to the statements made in the original
works on decoherence-free subspaces \cite{Lidar98}, where a
decoherence-free statistical operator was defined through annulment by the
Lindblad dissipator. Note, however, that here we identify the
decoherence-free subspaces in the system Liouville space, rather
than in its Hilbert space. This allows for the possibility that some
entangled system states ($\mathrm{Tr}\rho_S^2\neq\mathrm{Tr}\rho_S$)
could be resilient against decoherence, which is a potentially
useful feature that cannot be captured in the Hilbert space alone.

Theorem 1 gives us a straightforward,
general recipe for the classification of the decoherence-free
subspaces in the case of Markovian dynamics (\ref{eq:Markov final}).
What one needs to do is to construct the operator ${\Lambda}$
according to Eq. (\ref{eq:Lambda}), from the
microscopic interaction Hamiltonian and the environmental
preparation, solve its eigenproblem (in general numerically), and
investigate whether any of its null-spaces is invariant under
$\mathcal{L}_{\mathrm{eff}}.$ This is a simple, efficient way to
approximately determine where the information should be stored, and
should work well as long as the system is small enough to allow for
a full solution to the eigenproblem of $\Lambda$.

Moreover, the structure of the eigenspaces of ${\Lambda}$ enables us
to determine the directions of decoherence. For instance,
regardless of the value of $\tau$, we can still tell which states do
and which do not decohere, and calculate the relative values of the
decoherence rates for two given states. For fast switching in
nanoscale semiconductor devices, for example, we need rapid
coherence loss between the active region and leads, and we may
therefore opt to prepare the system in the subspace of $\Lambda$
corresponding to one of its largest eigenvalues.

\subsubsection{Identification of the steady state}

An important special case of a decoherence-free subspace is that of
a vector belonging to the intersection of $\mathcal
N(\mathcal{L}_{\mathrm{eff}})$  and $\mathcal N(\Lambda)$.

\textbf{Theorem 2.} \textit{A statistical operator belonging to $\mathcal
N({\mathcal{L}_{\mathrm{eff}}})\cap \mathcal N({\Lambda})$, the
intersection of the null-spaces $\mathcal
N(\mathcal{L}_{\mathrm{eff}})$  and $\mathcal N(\Lambda)$, is a
steady state for the evolution according to the map
(\ref{eq:Markov final}).}

\textit{Proof}. $\mathcal N({\mathcal{L}_{\mathrm{eff}}})\cap
\mathcal N({\Lambda})$ is the null space of the Markovian semigroup
generator. Consequently, any statistical operator prepared in $\mathcal
N({\mathcal{L}_{\mathrm{eff}}})\cap \mathcal N({\Lambda})$ remains
unchanged at all times, satisfying the definition of a steady state.
$\blacksquare$

By looking into the common null-subspace of both
$\mathcal{L}_{\mathrm{eff}}$ and $\Lambda$, one can narrow down the
set of potential steady states, which is important in many-body
transport calculations. In the case of a many-particle open
system, a full solution to the eigenproblem of $\Lambda$ may not be
tractable; however, identification of the common null-space of $\mathcal{L}_{\mathrm{eff}}$ and $\Lambda$
may be.

\subsubsection{A comment on the validity of Eq. (\ref{eq:Markov final})}

In general, whenever an efficient resetting mechanism can be defined for the environment, so that Eq. (\ref{eq:validity of Markov}) is satisfied,
(\ref{eq:Markov final}) should be applicable. However, it also appears that the simple equation (\ref{eq:Markov final}) may be used more broadly than specified by (\ref{eq:validity of Markov}).
Namely, on one of the few exactly solvable systems, the spin boson model with pure dephasing, which experiences Markovian evolution
in the long time limit regardless of the coupling strength, it can be shown (see Appendix \ref{sec:spinboson})
that one can define \textit{a mathematical coarse graining time $\tau$} that is shorter than any other timescale
in the coupled system and environment, so that coarse-grained evolution over
$\tau$ (\ref{eq:Markov final}) and the exact Markovian evolution coincide. So, it appears that not only does coarse
graining result in Markovian maps, but the converse might also be true:
it is possible that a given Markovian evolution can
be obtained by coarse graining of the short-time dynamics
if a suitable (ultrashort) mathematical coarse graining time is chosen. This statement
would, of course, be very difficult to
prove in general terms, but is interesting because it would mean that
all one needs to deduce the steady state for the
evolution of an open system is the information
on its short-time dynamics (\ref{eq:short time expansion of K}), which can in principle be done relatively straightforwardly and
from first principles (the microscopic interaction and the preparation of the environment).
Indeed, on an additional example of the Jaynes-Cummings Hamiltonian in the rotating wave approximation, which has been worked out in Appendix \ref{sec:Jaynes-Cummings},
it has been shown that by using map (\ref{eq:Markov final}) and the resulting criterion for the steady state (Theorem 2), relaxation towards the proper
equilibrium state has been obtained. So it appears that the applicability of Eq. (\ref{eq:Markov final})
may extend beyond the formal range of its validity (\ref{eq:validity of Markov}).

\section{A Two-Terminal Ballistic Nanostructure}\label{sec:Model}
In this Section, we consider a generic two-terminal nanostructure
under bias, and introduce a model interaction between the ballistic
active region and the contacts. This model should hold regardless of
whether the structure has resonances or not, as it is constructed to
mimic the source term in the single-particle density matrix
\cite{Jacoboni92,Rossi92,Brunetti89,Hohenester97,Ciancio04,Platero04}
and Wigner function
\cite{Kluksdahl89,Potz89,Frensley90,Bordone99,Biegel97,Jensen89,Grubin02,Shifren03,Jacoboni03,Nedjalkov04,Nedjalkov06}
formalisms, and preserve the continuity of current, state-by-state.
In Sec. \ref{sec:RTD}, the results are illustrated on a
one-dimensional two-barrier tunneling structure.

The left contact is the injector (source), biased negatively, while the right contact is the collector (drain). The contact-active region boundaries are at
$x_L$ (left) and $x_R$ (right), with $W=x_R-x_L$ being the active region width. We will assume that the active region
includes a large enough portion of the contacts (i.e., exceeding several Debye lengths) so that there is no doubt about the flat-band condition in the contacts. Also, $W$ should be large enough to reasonably ensure a quasicontinuum of wavevectors ($\Delta k=2\pi/W$) following the periodic boundary conditions. While sweeping the negative bias on the injector contact, we will assume
that it is done slowly (so that between two bias points the
system is allowed to relax) and in small increments (so that the potential profile inside the active region
does not change much between two bias points, and can be regarded
constant during each transient).

\begin{figure}
\includegraphics[width=8 cm]{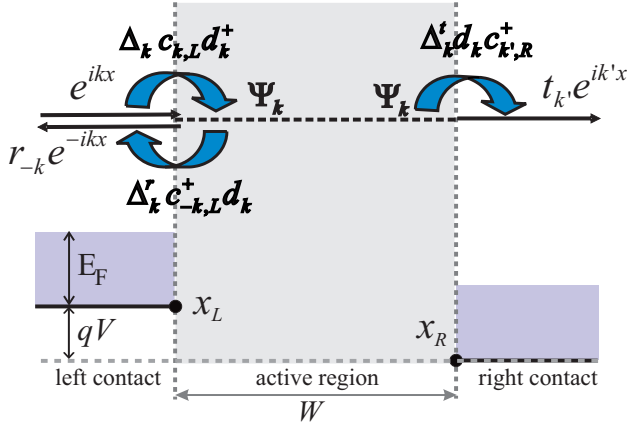}
\caption{(Color online) Schematic of the coupling between the active region of a generic two-terminal nanostructure and the contacts.
In case of ballistic injection through the open boundaries, a forward-propagating state $\Psi_k$ is coupled with the states
$\exp(\pm ikx)$ in the left contact (injected and reflected wave)
and the state $\exp(ik'x)$ in the right contact (transmitted wave) via a hopping model interaction (\ref{eq:Hplus}).}\label{fig:TwoTerminalSchematic}
\end{figure}

For every energy $\mathcal{E}_k$ above the bottom of the left contact, the active region's single particle Hamiltonian has two eigenfunctions,
a forward ($\Psi_k$) and a backward ($\Psi_{-k}$) propagating state, that can be found by (in general numerically) solving the single-particle Schr\"{o}dinger equation for a given potential profile in the active region. To keep the discussion as general as possible, we will not specify the details of how the active region actually looks (Fig. \ref{fig:TwoTerminalSchematic}) --
e.g., it can be a heterostructure, a \textit{pn} homojunction, or a MOSFET channel -- but we will require that the contact-active region open boundaries (at $x_L$ and $x_R$) are far enough from any junctions in the active region, so that the behavior of
$\Psi_{\pm k}$ near the junctions is already plane-wave like, i.e., that their general form near the injector is
$\Psi_k(x_L^+)=e^{ik{x_L^+}}+r_{-k,L}
e^{-ik{x_L^+}},\;\Psi_{-k}({x_L^+})=t_{-k,L}e^{-ik{x_L^+}},$ while near the collector
$\Psi_k({x_R^-})=t_{k',R}e^{ik'{x_R^-}},\;\Psi_{-k}({x_R^-})=e^{-ik'{x_R^-}}+r_{k',R}
e^{ik'{x_R^-}}$. Here,  where $t$'s and $r$'s are the transmission and reflection
amplitudes, while $k$ and $k'$ are the wavevectors that correspond to the same energy
$\mathcal{E}_k$, measured with respect to the conduction band bottoms in the left and right contacts, respectively
($k^2=k'^2-2mqV/\hbar^2=2m\mathcal{E}_k/\hbar^2$, where $-V$ has been applied to the left contact, and $q$ is the electron charge).

Associated with $\Psi_k$ ($\Psi_{-k}$) in the active region are the creation and
destruction operators $d_k^\dagger$ and $d_k$ ($d_{-k}^\dagger$ and
$d_{-k}$), so the active region many-body Hamiltonian is
\begin{equation}\label{eq:Hs}
\mathcal{H}_S=\sum_{k>0} \mathcal \omega_k (d_k^\dagger d_k
+d_{-k}^\dagger d_{-k}).
\end{equation}
Spin is disregarded, and $\omega_k=\mathcal{E}_k/\hbar$. In case of ballistic injection through the open boundaries, each state $\Psi_k$ is naturally coupled with the states
$\exp(\pm ikx)$ in the left contact (injected and reflected wave)
and the state $\exp(ik'x)$ in the right contact (transmitted wave).
For $\Psi_{-k}$, the coupling is
between $\exp(\pm ik'x)$ in the right and $\exp(-ikx)$ in the left contact.
To model this coupling via a hopping-type interaction, we can write quite generally (see Fig. \ref{fig:TwoTerminalSchematic})
\begin{eqnarray}\label{eq:Hplus}
\mathcal
H_{\mathrm{int}}^{+}&=&\sum_{k>0}\left(\Delta_k d_k^{\dagger}c_{k,L}+\Delta_k^r c_{-k,L}^{\dagger}d_k+\Delta_k^t c_{k',R}^{\dagger}d_k\right)+h.c.\nonumber\\
\end{eqnarray}
$c^\dagger_{\pm k,L}$ ($c_{\pm k,L}$) and $c^\dagger_{k',R}$ ($c_{k',R}$)
create (destroy) an electron with a wavevector $\pm k$ in the left and
$k'$ in the right contact, respectively. The hopping coefficients $\Delta_k$, $\Delta_k^r$ and $\Delta_k^t$ are the rates of injection, reflection, and transmission, respectively.
Therefore, they are proportional to the injected, reflected, and transmitted current for the state $\Psi_k$, i.e.,
\begin{equation}
\frac{\Delta_k^r}{\Delta_k}=\mathcal{R}_k,\;\; \frac{\Delta_k^t}{\Delta_k}=\mathcal{T}_k,
\end{equation}
where $\mathcal{R}_k$ and
$\mathcal{T}_k$ are the reflection and transmission coefficient at a
given energy.
The actual magnitude of $\Delta_k$ can be determined by requiring that $\Delta^t_k=\mathcal{T}_k\Delta_k$, the hopping rate from the active region into the right
contact, be the same as the current (per unit charge) carried through the active region by $\Psi_k$. This just means there is no more reflection
once the wave exits the active region and gets into the outgoing contact, and is usually referred to as the assumption of
reflectionless leads. \cite{Datta_book}) The current carried by $\Psi_k$ is given by the well-known quantum-mechanical relationship
\begin{eqnarray}\label{eq:j_k}
I_k&=&\frac{q\hbar}{m\|\Psi_k\|^{2}}\mathrm{Im}\left(\Psi_k^*\nabla\Psi_k\right)\nonumber\\
&=&\frac{q\hbar}{m}\frac{|t_{k',R}|^2
k'}{\left\|\Psi_k\right\|^2}=\frac{q\hbar k}{m}\frac{\mathcal{T}_k}{\left\|\Psi_{k}\right\|^2},
\end{eqnarray}
where we have used the form of $\Psi_k$ near the right contact
$t_{k',R}e^{ik'x}$, and ${\left\|\Psi_{k}\right\|^2}=\int_{0}^{W} dx
|\Psi_k(x)|^2$ is the norm squared of $\Psi_{k}$ over the active
region of width $W$. Since we require that
$I_k/q=\Delta^t_k=\mathcal{T}_k\Delta_k$, we find
\begin{equation}\label{eq:Delta_k}
\Delta_k=\frac{\hbar k}{m\left\|\Psi_{k}\right\|^2}.
\end{equation}

Finally, \begin{eqnarray}\label{eq:H+}
\mathcal
H_{\mathrm{int}}^{+}&=&\sum_{k>0}\Delta_k\left(d_k^{\dagger}c_{k,L}+\mathcal{R}_kc_{-k,L}^\dagger d_k+\mathcal{T}_k c_{k',R}^\dagger d_{k}\right)+h.c.,\nonumber\\
\end{eqnarray}
while the Hamiltonian for the backward propagating states
can be written in an analogous fashion, as
\begin{eqnarray}\label{Hminus}
\mathcal
H_{\mathrm{int}}^{\mathrm{-}}&=&\sum_{k>0}\Delta_{-k}\left(d_k^{\dagger}c_{-k',R}+\mathcal{R}_{-k}c_{k',R}^\dagger d_k+\mathcal{T}_{-k} c_{k',R}^\dagger d_{k}\right)+h.c.\nonumber\\
\end{eqnarray}
with $\Delta_{-k}=\frac{\hbar k'}{m\|\Psi_{-k}\|^2}$, and $\mathcal{T}_k=\mathcal{T}_{-k}$, $\mathcal{R}_k=\mathcal{R}_{-k}$.


When we put it all together, we have for the interaction Hamiltonian of
the active region with the left/right contact:
\begin{subequations}\label{H_L&H_R}
\begin{eqnarray}
\mathcal
H_{\mathrm{int}}^L&=&\sum_{k>0}\Delta_k\left\{\left(c_{k,L}^{\dagger}+\mathcal{R}_{k}c_{-k,L}^\dagger\right)d_k\right.\nonumber\\
&+&\left.d_k^\dagger\left(c_{k,L}+\mathcal{R}_{k}c_{-k,L}\right)\right\}\label{H_L}\\
&+&\Delta_{-k}\left\{\mathcal{T}_{k}c_{-k,L}^\dagger d_{-k}
+\mathcal{T}_{k}d_{-k}^\dagger c_{-k,L} \right\},\nonumber\\
\mathcal
H_{\mathrm{int}}^R&=&\sum_{k>0}\Delta_{-k}\left\{\left(c_{-k',R}^{\dagger}+\mathcal{R}_{k}c_{k',R}^\dagger\right)d_{-k}\right.\nonumber\\
&+&\left.d_{-k}^\dagger\left(c_{-k',R}+\mathcal{R}_{k}c_{k',R}\right)\right\}\label{H_R}\\
&+&\Delta_k\left\{\mathcal{T}_{k}c_{k',R}^\dagger d_k
+\mathcal{T}_{k}d_k^\dagger c_{k',R} \right\}.\nonumber
\end{eqnarray}
\end{subequations}

\subsection{Current-Carrying Contacts and the Approximation of a Memoryless Environment}
\label{sec:CurrentCarryingContactsMemoryless}

Now that we have the interaction Hamiltonians in place, we should
evaluate the matrix elements of the superoperator $\Lambda$, which
leads us to the questions how the approximation of a
memoryless environment is actually applied to contacts carrying
current, and how the expectation values of the interaction
Hamiltonian are to be calculated.

In general, as the current flows through the structure, we must
allow for different distributions of the forward and backward
propagating waves in the left and right contacts to ensure current
continuity. A simple and often employed approximation for the
steady-state distribution in the contacts carrying current $I$ is a
single-parameter drifted (or displaced) Fermi-Dirac distribution
\cite{Fischetti98,Fischetti99}
\begin{equation}\label{eq:drifted Fermi-Dirac distrib}
{f}^L_{\pm k}=\langle n_{\pm
k,L}\rangle=\frac{1}{\exp\left\{\frac{\hbar^2[(\pm
k-k_d)^2-k_F^2]}{2mk_BT}\right\}+1}.
\end{equation}
Here, $k_F$ is the Fermi wavevector and $k_d$ is the drift
wavevector, determined from the total current $I$ as
$k_d=mI/n_{1D}q\hbar$, where $n_{1D}$ is the 1D carrier density in
each contact (contacts are assumed identical). A drifted Fermi-Dirac
distribution, with the temperature equal to that of the lattice, is
often employed when we are interested in just the first two moments
of the distribution function (i.e., maintaining charge neutrality
and ensuring current continuity). Additionally, if needed,
information on the electron heating can be incorporated in this
distribution by allowing for a discrepancy between the electronic
and lattice temperatures (we will neglect electron heating here).
Detailed ensemble Monte Carlo - molecular dynamics simulations of
carrier transport in highly doped ($>10^{17}\;\mathrm{cm}^3$) bulk
semiconductors, in which electron-electron scattering is the most
efficient energy relaxation mechanism, have shown to produce
distributions very close to the drifted Fermi-Dirac distribution
(\ref{eq:drifted Fermi-Dirac distrib}),
\cite{Lugli85,Lugli86,Kriman90} which is generally accepted as a
decent approximation for these systems. Here, we will also adopt
(\ref{eq:drifted Fermi-Dirac distrib}) for the distribution of
carriers in the current-carrying contacts, and it is reasonable if
the (one-dimensional) contacts are longer than $\sqrt{D\tau}$, where
$D$ is the diffusion constant (otherwise, the distribution function
in them may not be thermalized \cite{Pohtier97,Pierre00}).

Now, the question arises what happens if we try to sweep the
voltage.  We have mentioned before that the voltage is to be swept
slowly (enough time between two bias points for the system to relax)
and in small increments (so that we can consider the barrier as
having a constant profile during each transient). The latter is
crucial for the implementation of the approximation of a memoryless
environment. Suppose that, at a bias $V$, a steady-state current $I$
is flowing through the structure. If we increase the bias to $\Delta
V$ at $t=0$, where $\Delta V$ is very small, within the first
$t=\tau$, the current is virtually unchanged -- it takes the current
a much longer time $\tau_{AR}\gg\tau$ to change significantly (AR
stands for "active region"; once we have had a chance to complete
the calculation, we will see that $\tau_{AR}$ will be equal to
$1/\lambda\tau$, where $\lambda$ is a relevant eigenvalue of
$\Lambda$). Therefore, after $\tau$, the contact carriers have
redistributed themselves to the old distribution function that they
had at $V$. Basically, the contact carriers as redistributing
themselves to Eq. (\ref{eq:drifted Fermi-Dirac distrib}) determined
by the (virtually) instantaneous current level at each $\tau$; the
current, however, changes very little during each $\tau$. By the
time the current has saturated ($\sim\tau_{AR}$), the contact
carriers have had a chance to get redistributed many times; however,
if the total voltage increase $\Delta V$ is very small, the total
current increase during the full transient will also be small, so we
can say that during the whole transient the distribution functions
of the forward and backward propagating states have been resetting
to nearly the same distribution, approximately the average of
$f_{\pm k}^{L,R}$ over the interval $[V,V+\Delta V]$. Clearly, as
the voltage sweep increment $\Delta V\rightarrow 0$, we can say that
during a transient the contacts redistribute to $f_{\pm k}^{L,R}$
(\ref{eq:drifted Fermi-Dirac distrib}) at $V$.

Evaluation of $k_d$ that enters the contact distribution functions
at a given voltage must be done self-consistently: starting with a
guess for $k_d$ at a given voltage, steady-state distributions and
current are evaluated (as detailed in the next section). The
obtained current is then used to recalculate $k_d$, and the process
is repeated until a satisfactory level of convergence is achieved.
(Of course, the initial guess for $k_d$ at any voltage can be
$k_d=0$, but for faster convergence it is better to start with the
$k_d$ found for the preceding voltage.)

\subsection{Markovian Relaxation for a Two-Terminal Nanostructure.
Steady-State Distributions and Current}\label{sec:Lambda for model and steady state distributions}

Since the interaction Hamiltonians (\ref{H_L&H_R}) are linear in the
contact creation and destruction operators, and we can approximate
that each contact snaps back to a "drifted" grand-canonical
statistical operator, we have $\langle \mathcal
H_{\mathrm{int}}^{L/R}\rangle =0$. This means that
${\mathcal{L}}_S=\mathcal{L}_{\mathrm{eff}}$, and also leaves us
with only the first three terms in Eq. (\ref{eq:Lambda}) for
$\Lambda$ to calculate. One can show that $\Lambda=\Lambda^L
+\Lambda^R$, where
\begin{eqnarray}\label{Lambda_L,R}
(\Lambda^{L/R})^{\alpha,\beta}_{\alpha ',\beta
'}&=&\frac{1}{2}\left(\langle \left({\mathcal
H}_{\mathrm{int}}^{L/R}\right)^2\rangle^{\alpha}_{\alpha '}\delta^{\beta
'}_{\beta }+\langle \left({\mathcal
H}_{\mathrm{int}}^{L/R}\right)^2\rangle^{\beta
'}_{\beta }\delta^{\alpha}_{\alpha '}\right)\nonumber\\
&-&\sum_{i,j}\rho_{L,R}^i
({\mathcal H}_{\mathrm{int}}^{L/R})^{j\alpha}_{i\alpha '}({\mathcal
H}_{\mathrm{int}}^{L/R})^{i\beta '}_{j\beta }.
\end{eqnarray}
The first and the second term in Equation (\ref{Lambda_L,R}) give a
general contribution of the form
$\Lambda^{\alpha\beta}_{\alpha\beta}$, since
\begin{eqnarray}\label{H^2}
\langle \left({\mathcal
H}_{\mathrm{int}}^{L}\right)^2\rangle
&=&\sum_{k>0}\Delta^2_{k}\left\{\left(\langle n_{k,L}\rangle
+\mathcal{R}_{k}^2\langle n_{-k,L}\rangle\right)d_k
d_k^\dagger\nonumber\right.\\
&+&\left.\left[(1-\langle
n_{k,L}\rangle)+\mathcal{R}_{k}^2(1-\langle n_{k,L}\rangle)\right]
d_k^\dagger d_k\right\}\nonumber\\
&+&\Delta^2_{-k}\left\{\mathcal{T}_{k}^2\langle n_{-k,L} \rangle
d_{-k}
d_{-k}^\dagger\right.\nonumber\\
&+&\left.\mathcal{T}_{k}^2\left(1-\langle n_{-k,L}\rangle\right)
d_{-k}^\dagger d_{-k}\right\}\nonumber\\
&=&\sum_{k>0}\Delta^2_{k}\left\{\left(f^L_k
+\mathcal{R}_{k}^2f^L_{-k}\right)d_k
d_k^\dagger\nonumber\right.\\
&+&\left.\left[(1-f^L_{k})+\mathcal{R}_{k}^2(1-f^L_{-k})\right]
d_k^\dagger d_k\right\}\nonumber\\
&+&\Delta^2_{-k}\left\{\mathcal{T}_{k}^2f^L_{-k}  d_{-k}
d_{-k}^\dagger\right.\nonumber\\
&+&\left.\mathcal{T}_{k}^2\left(1-f^L_{-k}\right)
d_{-k}^\dagger d_{-k}\right\}\nonumber\\
\end{eqnarray}
preserves the filling of states. We have used $\langle n_{\pm k,L}
\rangle=f^L_{\pm k} $, where $f^L_{\pm k}$ is the drifted
Fermi-Dirac distribution function in the left contact
(\ref{eq:drifted Fermi-Dirac distrib}).

In contrast, the third term in (\ref{Lambda_L,R})
\begin{eqnarray}\label{Lambda RTD occupation changing term}
&{}&\sum_{i,j}\rho_{L}^i ({\mathcal
H}_{\mathrm{int}}^L)^{j\alpha}_{i\alpha '}({\mathcal
H}_{\mathrm{int}}^L)^{i\beta '}_{j\beta }=\nonumber\\
&=&\sum_{k>0}\Delta^2_{k}
\left\{\left[(1-f^{L}_k)+\mathcal{R}_k^2(1-f^{L}_{-k})\right](d_k)^\alpha_{\alpha
'}(d_k^\dagger)^{\beta '}_{\beta}\right.\nonumber\\
&+&\left.\left(f^{L}_k+\mathcal{R}_k^2f^{L}_{-k}\right)(d_k^\dagger)^\alpha_{\alpha
'}(d_k)^{\beta
'}_{\beta}\right\}\\
&+&\Delta^2_{-k}\left\{{\mathcal{T}_k^2}(1-f^{L}_{-k})
(d_{-k})^{\alpha}_{\alpha
'}(d_{-k}^\dagger)^{\beta'}_{\beta}\right.\nonumber\\
&+&\left.{\mathcal{T}_k^2} f^L_{-k}
(d_{-k}^\dagger)^{\alpha}_{\alpha
'}(d_{-k})^{\beta'}_{\beta}\right\}\nonumber
\end{eqnarray}
gives a contribution of the form
$\Lambda^{\alpha\alpha}_{\beta\beta}$.

Each term in $\Lambda$ is a sum of independent contributions over
individual modes [$\Lambda=\sum_{k} \Lambda_k$] that attack only
single-particle states with a given $k$. The same holds for
$\mathcal {\mathcal{L}}_S$. Consequently, in reality we have a
multitude of two-level problems (see Appendices \ref{sec:spinboson} and \ref{sec:Jaynes-Cummings}),
one for each state $\Psi_{k}$,
where the two levels are a particle being in $\Psi_{k}$ ("+") and a
particle being absent from $\Psi_{k}$ ("-"). Each such 2-level
problem is cast on its own 4-dimensional Liouville space, with
$\rho_k=\left(\rho^{++}_k,\rho^{+-}_k,\rho^{-+}_k,\rho^{--}_k\right)^\mathrm{T}$
being the reduced statistical operator that describes the occupation of
$\Psi_{k}$. According to (\ref{eq:Markov final}),
\begin{equation}
\frac{d\rho_k}{dt}=[-i\mathcal
{\mathcal{L}}_{S,k}-\Lambda_k\tau]\rho_k,
\end{equation}
where
\begin{subequations}
\begin{eqnarray}
{\mathcal{L}}_{S,k}&=&\left[\begin{array}{cccc}
0 & 0 & 0 & 0 \\
0 & 2\omega_k & 0 & 0 \\
0 & 0 & -2\omega_k & 0 \\
0 & 0 & 0 & 0 \end{array}\right],\\
{\Lambda}_k&=&\left[\begin{array}{cccc}
A_k & 0 & 0 & -B_k\\
0 & C_k & 0 & 0 \\
0 & 0 & C_k& 0 \\
-A_k & 0 & 0 & B_k\end{array}\right],
\end{eqnarray}
\end{subequations}
and
$A_k=\Delta^2_{k}\{(1-f^{L}_k)+{\mathcal{R}_k^2}(1-f^{L}_{-k})+{\mathcal{T}_k^2}(1-f^R_{k'})\}$,
$B_k=\Delta^2_{k}\{f^{L}_k+{\mathcal{R}_k^2}f^{L}_{-k}+{\mathcal{T}_k^2}f^R_{k'}\}$,
and
$C_k=(A_k+B_k)/2=\Delta^2_{k}\left(1+{\mathcal{R}_k^2}+{\mathcal{T}_k^2}\right)/2$.
The rows/columns are ordered as
$1=\left|+\right\rangle\left<+\right|,
2=\left|+\right\rangle\left<-\right|,
3=\left|-\right\rangle\left<+\right|,
4=\left|-\right\rangle\left<-\right|$. The diagonal elements in
$\Lambda_k$ originate from the terms of the form
$\Lambda^{\alpha\beta}_{\alpha\beta}$, calculated using Eq.
(\ref{H^2}), while the off-diagonal ones originate from
$\Lambda^{\alpha\alpha}_{\beta\beta}$ (\ref{Lambda RTD occupation
changing term}). Strictly speaking, the time evolution above is
valid if (\ref{eq:validity of Markov}) is satisfied, which in this
case implies $\Delta^2_{\pm k}\tau^2\ll\min\{1,\omega_k\tau\}$.
After approximating $\|\Psi_k\|^2\approx W$, we obtain the condition
$(v\tau/W)^2\ll\min\{1,\omega_k\tau\}$, where $v=\hbar k/m$. For
typical values of $W=100$nm, $v\leq v_F=10^5$ m/s, and $m=0.067m_0$
appropriate for GaAs, both equations will be satisfied for $\tau\ll
1$ ps.

Clearly, off-diagonal elements $\rho_k^{+-}$ and $\rho_k^{-+}$
decay as $\exp{(\mp i2\omega_k-\tau C_k)t}$ and are zero in the
steady state. The two equations for $\rho_{k}^{++}=f_k(t)$ and
$\rho_{k}^{--}=1-f_k(t)$ are actually one and the same, and either
one yields
\begin{eqnarray}\label{eq:f time evolution}
\frac{df_k(t)}{dt}&=&-\tau A_kf_k(t)+\tau B_k[1-f_k(t)]\nonumber\\
&=&-\tau(A_k+B_k)f_k(t)+\tau B_k ,
\end{eqnarray}
where $f_k$ is the distribution function for the active region. In
the steady state, we have $f_k^\infty=\frac{B_k}{A_k+B_k}$ (for
$-k$, by analogy), so finally
\begin{subequations}\label{eq:f's}
\begin{eqnarray}
f_k^\infty&=&\frac{f^L_{k}+{\mathcal{R}_k^2}f^L_{-k}+
{\mathcal{T}_k^2}f^R_{k'}}{1+{\mathcal{R}_k^2}+{\mathcal{T}_k^2}^2},\quad\\
f_{-k}^\infty&=&\frac{f^R_{-k'}+{\mathcal{R}_k^2}f^R_{k'}
+{\mathcal{T}_k^2}f^L_{-k}}{1+{\mathcal{R}_k^2}+{\mathcal{T}_k^2}}.
\end{eqnarray}
\end{subequations}

Note that there is no dependence of the steady-state distributions
on $\Delta_{k}$, the hopping interaction strength, or the
coarse-graining time $\tau$. $f_{\pm k}^\infty$ obviously differ
from the contact distributions (see discussion in the next section).
The discontinuity of the distribution functions across each open
boundary is a price to pay to conserve the flux across it, the same
as in the heuristic treatment of carrier injection in the density
matrix, Wigner function, and Pauli equation formalisms (see the
discussion on p. 4907 of Ref. \onlinecite{Fischetti99}).

The steady-state current (per spin orientation) can be calculated as
\begin{eqnarray}
I^{\infty}=\sum_{k>0} f^{\infty}_k I_k+f^{\infty}_{-k}I_{-k},
\end{eqnarray}
where $I_{k}=q\hbar k\mathcal{T}_k/m\|\Psi_k\|^2$ and $I_{-k}=q\hbar
k'\mathcal{T}_k/m\|\Psi_{-k}\|^2$ (\ref{eq:j_k}). $I_{\pm k}$ are
each constant across the active region and given by . The total
current carried by the forward propagating states (per spin
orientation) is
\begin{subequations}
\begin{eqnarray}
I_{+}^{\infty}&=&\frac{q\hbar}{m}\sum_{k>0}f^{\infty}_k\frac{k\mathcal{T}_k}{\left\|\Psi_k\right\|^2}\nonumber\\
&=&\frac{q\hbar}{m}\frac{W}{2\pi}\int_0^\infty k\;dk \;f^{\infty}_k \frac{\mathcal{T}_k }{\left\|\Psi_k\right\|^2}\\
&=&\frac{qW}{h}\int_0^\infty\,d\mathcal{E}_k\;f^{\infty}_{k}\frac{\mathcal T_k}{\left\|\Psi_{k}\right\|^2},\nonumber
\end{eqnarray}
where we have used $k\,dk=m\,d\mathcal E_k/\hbar^2$ and $\Delta k\approx W/2\pi$. Similarly, the current component (per spin) carried by
the backward propagating states is
\begin{eqnarray}
I_{-}^{\infty}
&=&-\frac{qW}{h}\int_0^\infty\,d\mathcal{E}_k\;f^{\infty}_{-k}\frac{\mathcal{T}_k}{\left\|\Psi_{-k}\right\|^2},
\end{eqnarray}
\end{subequations}
so the total current (per spin orientation) can be found as
\begin{equation}\label{eq:j total exact}
I^{\infty}=\frac{qW}{h}\int_0^\infty\,d\mathcal{E}_k\;\left(\frac{f^{\infty}_{k}}{\left\|\Psi_{k}\right\|^2}-
\frac{f^{\infty}_{-k}}{\left\|\Psi_{-k}\right\|^2}\right)\mathcal{T}_k.
\end{equation}
This expression is parameter-free,  because
$\left\|\Psi_{\pm k}\right\|^2$ in the denominator scale with $
W$.

\subsection{Relationship to the Landauer-B\"{u}ttiker
formalism}\label{sec:LB comparison}

A natural question emerging at this point is how the current
(\ref{eq:j total exact}) relates to that predicted by the
Landauer-B\"{u}ttiker (LB) formalism
\cite{Landauer57,Landauer70,Buttiker85,Buttiker86} (comprehensive
reviews of the LB formalism can be found, for instance, in Refs.
\onlinecite{Stone88} and \onlinecite{Blatner00}, as well as in many
textbooks \cite{FerryGoodnick,Datta_book}). The one-channel variant
of the current formula is referred to as the Landauer formula,
\begin{equation}\label{eq:j Landauer}
I^{\infty}_{\mathrm{Lan}}=\frac{q}{h}\int_0^\infty\,d\mathcal{E}_k\;\left[\bar{f}^{L}(\mathcal{E}_{k})-\bar{f}^R (\mathcal{E}_{k})\right]\mathcal{T}_k,
\end{equation}
where $\bar{f}^{L}(\mathcal{E}_{k})$ and
$\bar{f}^R(\mathcal{E}_{k})$ are the equilibrium distributions in
the left and right reservoirs. \cite{Blatner00} Generalization to
multiple channels is due to B\"{u}ttiker.
\cite{Buttiker85,Buttiker86,Buttiker86_1,Buttiker88}

Both the LB approach and the approach presented here focus on
maintaining the carrier flux through the open boundaries between the
active region and the contacts. There is one major difference,
however. In the LB approach, what is known are the distributions of
the states entering the active region (in our notation, $f^L_k$ and
$f^R_{-k'}$); nothing is specified about the distributions of the
states going out of the structure ($f^L_{-k}$ and $f^R_{k'}$), as
they can be calculated by using the transfer matrix (a nice
exposition of this issue can be found in Refs.
\onlinecite{Buttiker92} and \onlinecite{Blatner00}). In contrast, in
the approach presented here, we need the information on
\textit{both} the incoming ($f^L_k$ and $f^R_{-k'}$) and the
outgoing states ($f^L_{-k}$ and $f^R_{k'}$) in the contacts in order
to calculate the distributions of the forward and backward
propagating states ($f^\infty_k$ and $f^\infty_{-k}$) in the active
region. The reason is that \textit{the information about the
outgoing distributions, supplied by the transfer matrix, is
destroyed in the contacts, where the inelastic scattering very
rapidly redistributes carriers.}

Our model for the inelastic current-carrying contacts can actually
be considered as complementary to the well-known model of voltage
probes. \cite{Buttiker86_1,Beenakker92,Texier00,Forster07} On
average, a voltage probe carries no current. Due to inelastic
scattering, the distribution function in a voltage probe is reset to
the equilibrium one on timescales much shorter than the response
time limited by the active region ($\tau_{AR}$). In contrast, there
is no voltage drop over a current-carrying contact (conduction band
bottom is flat), while the average current carried by it is
generally nonzero. Due to inelastic electron-electron scattering,
the distribution function in the current-carrying contact is reset
to a displaced Fermi-Dirac distribution on timescales much shorter
than $\tau_{AR}$.

\subsection{Example: A Double-Barrier Tunneling
Structure}\label{sec:RTD}

We illustrate the results of Sec. \ref{sec:Lambda for model and steady state distributions}
on a one-dimensional, double-barrier tunneling structure, formed on a
quantum wire in which only one subband is populated. The Fermi level
is at 5 meV with respect to the subband bottom. The well width
is 15 nm, the barrier thickness is 25 nm, and the barrier height is 15
meV. These result in one bound state at about 6.84 meV when no bias is
applied. The goal is to calculate the nonequilibrium steady-state
distribution functions specified by Eq. (\ref{eq:f's}) under any
given bias $V$, and use this information to construct the I--V
curve. For simplicity, in this calculation the voltage is assumed to
drop linearly across the well and barriers, but in general, Eqs.
(\ref{eq:f's}) need to be coupled with a Poisson and a
Schr\"{o}dinger solver to obtain a realistic potential profile and
charge distribution.

\begin{figure}
\center\includegraphics[width=8 cm]{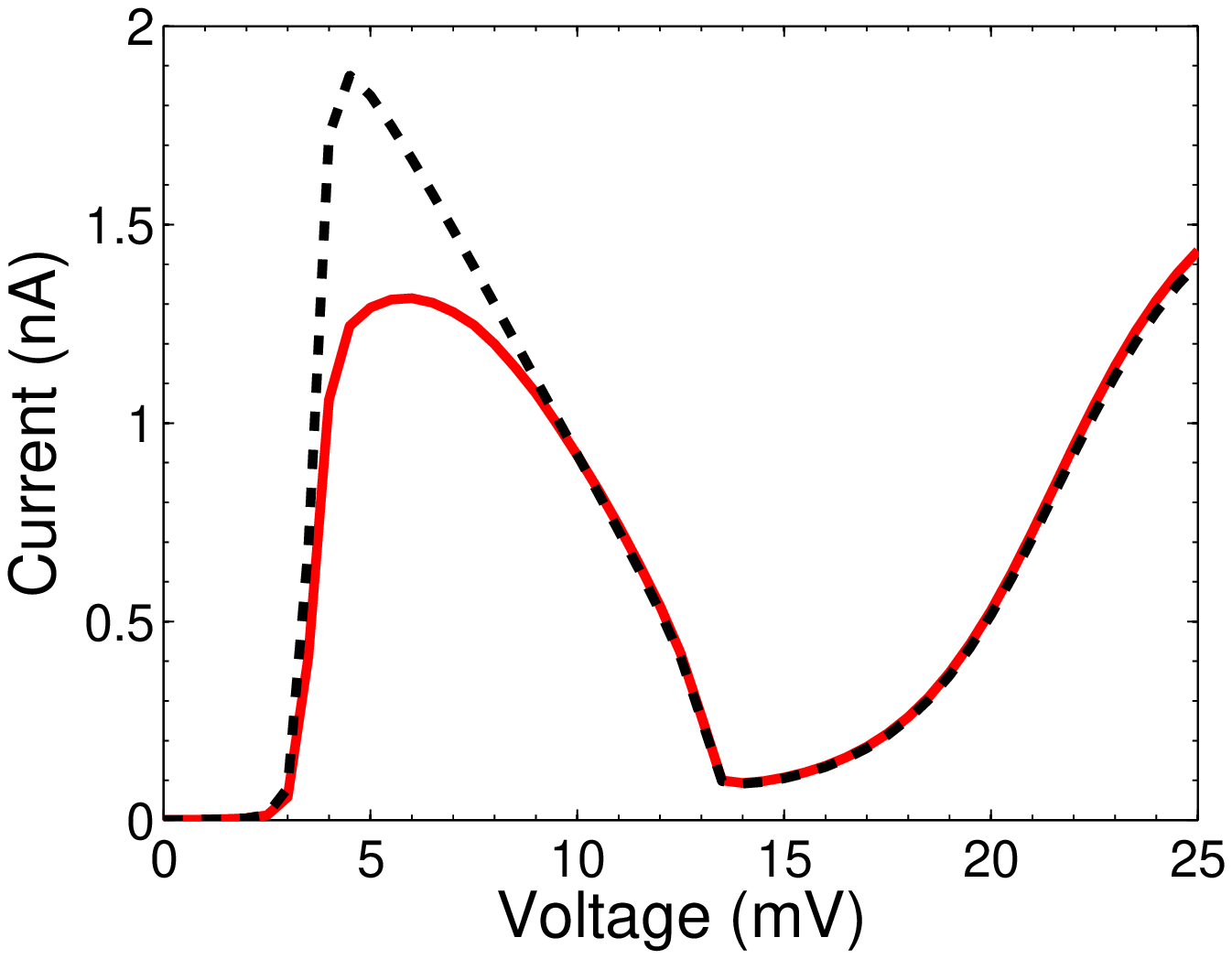}
\caption{\label{fig:I_Vcurve} (Color online) I--V curve for the
double-barrier tunneling structure, according to the expression
(\ref{eq:j total exact}) (solid curve) and the Landauer formula
(\ref{eq:j Landauer}) (dashed curve) at 1 K.}
\center\includegraphics[width=8 cm]{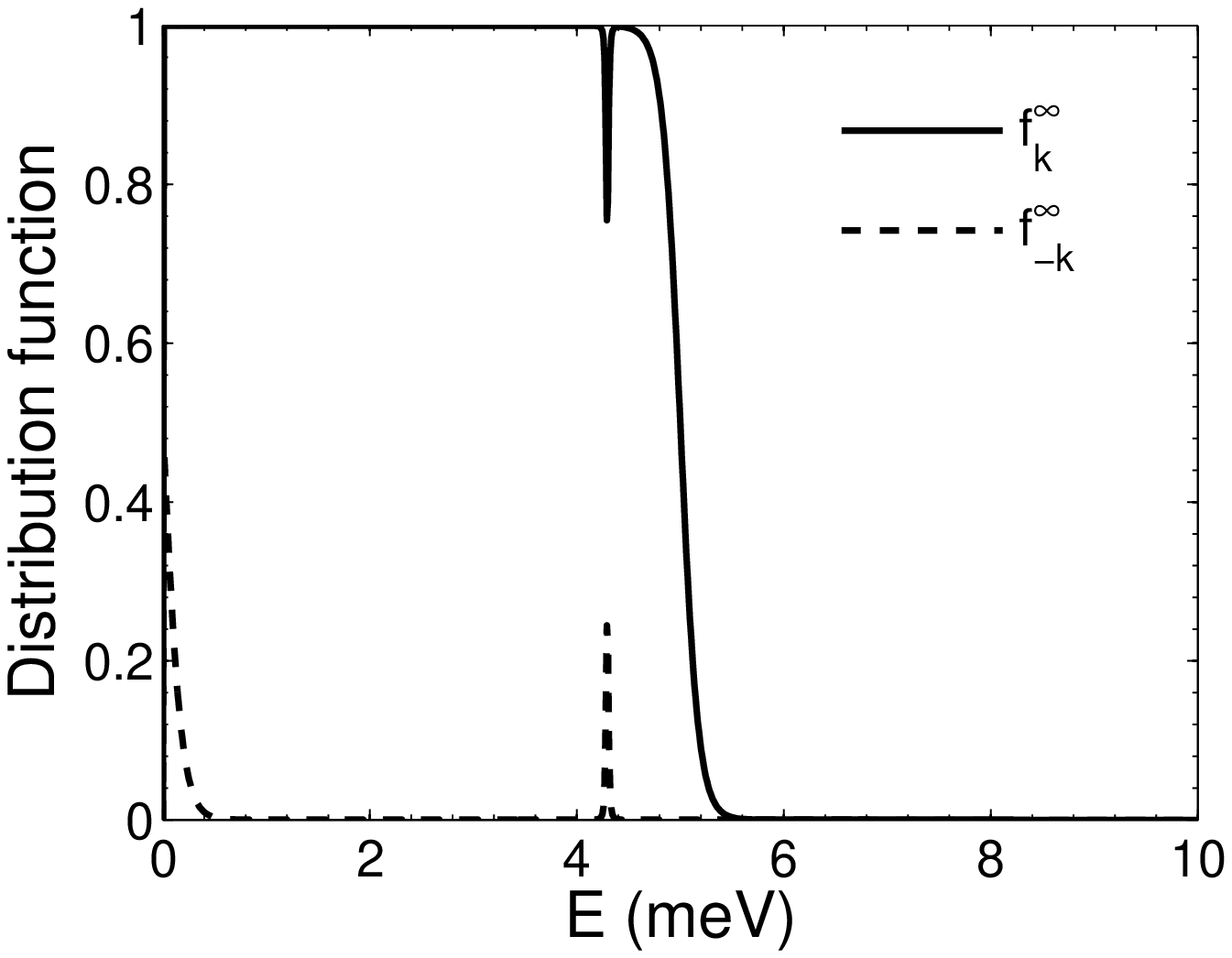}
\caption{\label{fig:Distribution and Transmission} Steady-state
distribution functions (\ref{eq:f's}) of the forward
($f^{\infty}_{k}$) and backward ($f^{\infty}_{-k}$) propagating
states, at the peak voltage from Fig. \ref{fig:I_Vcurve} (5 meV) and
1 K. Energy is measured with respect to the bottom of the injector
contact (Fig. \ref{fig:TwoTerminalSchematic}). Significant
deviations from the Fermi-Dirac equilibrium distributions in the
contacts coincide with the peak in transmission.}
\end{figure}

Figure \ref{fig:I_Vcurve} shows the I-V curve of the double-barrier
tunneling structure, as calculated according to the expression
(\ref{eq:j total exact}) and the Landauer formula (\ref{eq:j
Landauer}). In the voltage range depicted, the current flowing
through the structure is so low  that the equilibrium distribution
functions in the contacts ($\bar{f}^L(\mathcal{E}_k)$ and
$\bar{f}^R(\mathcal{E}_k)$) and the drifted Fermi-Dirac
distributions (\ref{eq:drifted Fermi-Dirac distrib}), with $k_d$
determined self-consistently, are extremely close to one another,
and give almost identical $f^\infty_{\pm k}$ (\ref{eq:f's}) and the
values for the current (\ref{eq:j total exact}). The difference
between the curves obtained by using the equilibrium contact
distributions and the drifted Fermi-Dirac is barely visible within
the voltage range presented (the maximal difference between the
currents obtained these two ways is $\approx 10^{-11} A$).

Both (\ref{eq:j total exact}) and (\ref{eq:j Landauer}) describe
ballistic transport, so no crossing of the curves typical for the
inclusion of inelastic scattering should be expected (inelastic
scattering causes the peak to lower and the valley to rise, so the
curves cross \cite{Nedjalkov04}). Both curves in Fig.
\ref{fig:I_Vcurve} properly display the resonant features, but the
Landauer formula (\ref{eq:j Landauer}) predicts a higher peak
current than (\ref{eq:j total exact}). The reason is that
$f^\infty_{\pm k}$, used in (\ref{eq:j total exact}),  coincide with
the contact (nearly equilibrium) distribution functions only if the
transmission is not high. Near a transmission peak, significant
deviations of $f^\infty_{\pm k}$ (\ref{eq:f's}) from the contact
distribution functions occur, as shown in Fig. \ref{fig:Distribution
and Transmission} for the peak voltage from Fig. \ref{fig:I_Vcurve},
and lead to the lowering of the current observed in Fig.
\ref{fig:I_Vcurve}.

\section{Summary and Concluding Remarks}\label{sec:Conclusion}

In this paper, a simple theoretical description of the
contact-induced decoherence in two-terminal nanostructures was
provided within the framework of the open systems theory. The model
active region -- contact interaction was introduced to ensure proper
carrier injection from the contacts. The steady-state statistical
operator of the active region was calculated by relying on the
Markovian map derived through coarse graining of the exact short
time dynamics over the energy relaxation time of the bulk-like
contacts. The ballistic-limit, steady state distribution functions
of the forward and backward propagating states for a generic
two-terminal nanostructure have been derived. The approach was
illustrated on the example of a double-barrier tunneling structure,
where an I-V curve that shows all the prominent resonant features
was obtained. The relationship between the present approach and the
Landauer-B\"{u}ttiker formalism was addressed.

The inclusion of scattering within the active region would alter the
form of $\mathcal{L}_{\mathrm{eff}}$, while scattering between the
active region and the contacts (e.g. phonon-assisted tunneling)
would essentially alter $\Lambda$. Equations (\ref{eq:f's}) are the
ballistic limit of the active region's nonequilibrium steady-state
distributions, and are a better starting point for transport
calculations with scattering than the equilibrium distributions: for
instance, the single-particle density matrix
$\rho^{(1)}(k_1,k_2)=\mathrm{Tr}_S(d^\dagger_{k_2}d_{k_1}\rho_S)$ in
the ballistic limit is obviously diagonal, so to include scattering
within the active region, one simple way would be to follow the
single-particle density matrix formalism,
\cite{Jacoboni92,Rossi92,Brunetti89,Hohenester97,Ciancio04,Platero04}
with the diagonal $\rho^{(1)}(k,k)$ specified by (\ref{eq:f's}) as
the ballistic limit. [Clearly, $\rho^{(1)}$ would no longer be
diagonal in $k$ once scattering is included.] Scattering due to
phonons within the active region is generally amenable to a
weak-coupling approximation, so it can be treated as a perturbation
within the Born approximation. To treat phonon-assisted injection
from the contacts, the contact many-body Hilbert space can be
augmented to formally include a tensor product of the contact and
the phonon Hilbert spaces, \cite{Fischetti98,Fischetti99} but again
a simpler perturbative treatment may be enough. As for the treatment
of electron-electron scattering, $\Lambda$ is in the form that
allows for its inclusion between the active region and the contacts,
but this is likely to be a difficult technical issue.

Finally, an important feature of the present approach is that it can be, at least in principle,
extended to arbitrarily short timescales by forgoing the coarse-graining procedure, so non-Markovian effects can be observed.
However, since the coarse-graining procedure phenomenologically accounts for the efficient electron-electron interaction in the contacts, without it we would be
required to explicitly include this interaction in the contact Hamiltonian, which will require certain modifications to the present approach.

\section{Acknowledgement} The author thanks D. K. Ferry, J. P. Bird, W. P\"{o}tz, and J. R. Barker for
helpful discussions. This work has been supported by the NSF, award
ECCS-0547415.

\appendix

\section{Uncorrelated initial state and the existence of a subdynamics}\label{appendix:uncorr}

In this Appendix, for an uncorrelated initial state of the form
$\rho (0)=\rho_E\otimes\rho_S(0)$, we will explicitly show that
$\rho_2(0)$, the component of $\rho(0)$ belonging to the zero
eigenspace of $\overline P$, can be written in terms of $\rho_S(0)$
via equation  (\ref{rho2vsrhoS}), repeated here
\begin{equation}\label{rho2vsrhoS:appendix}
\rho_2(0)=\mathcal M\rho_1(0)=d_E^{-\frac{1}{2}}\mathcal M\rho_S(0).
\end{equation}
Together with  (\ref{rhoSbasic}), this equation proves that a
subdynamics exists. We will explicitly derive the mapping $\mathcal
M$ that is uniquely fixed by $\rho_E$.

\subsection{Eigenbasis of $\overline P$}
Let us first remind ourselves of the structure of the eigenspaces of
$\overline P$. Its unit eigenspace is $d_S^2$-dimensional, spanned
by vectors of the form
\begin{equation}\label{alpha beta overline}
\left|\overline{\alpha\beta}\right\rangle=d_E^{-\frac{1}{2}}\sum_{i=1}^{d_E}
\left|{i\alpha,i\beta}\right\rangle .
\end{equation}
This form holds regardless of the environmental basis chosen, which
is in agreement with the fact that the uniform environmental statistical operator $\overline \rho_E$ (the one that induces $\overline P$) is a
scalar matrix, i.e., diagonal in any environmental basis. In the
zero eigenspace of $\overline P$, for any choice of the
environmental basis, we can identify two subspaces:

\noindent 1) A subspace spanned by vectors of the form
$\left|i\alpha,j\beta\right\rangle$, with $i\neq j$. This subspace
is $d_S^2d_E(d_E-1)$-dimensional.

\noindent 2) A subspace spanned by linear combinations of
$\left|{i\alpha,i\beta}\right\rangle$, which are orthogonal to all
$\left|\overline{\alpha\beta}\right\rangle$. These are given by
\begin{equation}
\left|b_{i,\alpha\beta}\right\rangle
=\sqrt{\frac{d_E+1-i}{d_E-i}}\left(\left|{i\alpha,i\beta}\right\rangle-
\frac{\sum_{j=i}^{d_E}\left|{j\alpha,j\beta}\right\rangle}{d_E+1-i}\right)
\end{equation}
for every pair $\alpha ,\beta$ and for $i=1,\dots , d_E -1$. This
subspace is $d_S^2(d_E-1)$-dimensional. Note how the coefficients in
the linear combinations do not depend on $\alpha ,\beta$.

\begin{figure}
\center\includegraphics[width=8cm]{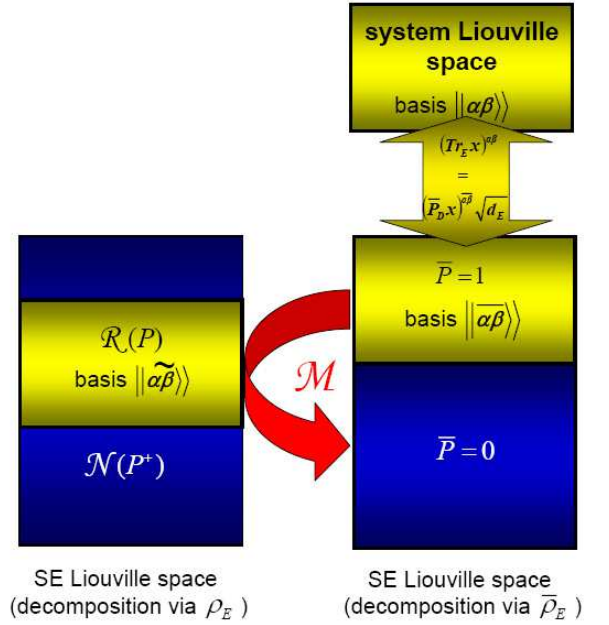}
\caption{\label{Fig 4.} (Color online) Two decompositions of the total $SE$
Liouville space: the first one (right) is into the eigenspaces of
$\overline P$, induced by the uniform statistical operator $\overline
\rho_E$. (System Liouville space and $\overline P$'s unit eigenspace
are equivalent.) The other decomposition is into $\mathcal{R}(P)$,
the range of the projector $P$ induced by the initial environmental
statistical operator $\rho_E$, and $\mathcal{N}(P^\dagger)$, the null
space of $P^\dagger$ and orthocomplement to $\mathcal{R}(P)$. These
two decompositions enable us to construct the mapping $\mathcal M$
used in the reduced dynamics (\ref{eq:dot rho_S uncorrelated}).}
\end{figure}

\subsection{Range (Image Space) of  $P$ and Null Space of $P^\dagger$}
Now let us get back to the initially uncorrelated state of the form
$\rho (0)=\rho_E\otimes \rho_S(0)$, and choose the eigenbasis of
$\rho_E$ as the environmental basis $|j\rangle ,j=1,\dots d_E$.
$\rho_E$, the initial environmental statistical operator from
$\rho(0)=\rho_E\otimes\rho_S(0)$ induces its own projection operator
$P$, so that for any vector $\mu$ from the $SE$ Liouville space we
can write
\begin{eqnarray}\label{Pmu}
P\mu&=&\rho_E\otimes\left(\mathrm{Tr}_E\mu\right)\nonumber\\
&=&\sum_{i,\alpha ',\beta '}\rho_E^i \left(\sum_k \mu^{k\alpha
',k\beta '}\right)\left|i\alpha,i\beta\right\rangle
\end{eqnarray}
The Latin indices $i,k=1,\dots d_E$ count environmental states,
while the Greek ones $\alpha,\beta=1,\dots d_S$ count the system
states. The essence of the following proof is to write any $P\mu$ in
terms of the eigenvectors of $\overline P$, and then, since $P\rho
(0)=\rho (0)$, draw important conclusions about its components
$\rho_2(0)$ and $\rho_S(0)$.

$P$ is not Hermitian or diagonalizable. We can, however, still speak
of its range (space of images) $\mathcal{R}(P)$, to which $\rho (0)$
belongs becaue $P\rho (0)=\rho (0)$. The orthocomplement to
$\mathcal{R}(P)$ is $\mathcal{N}(P^\dagger)$, the null space of the
adjoint operator $P^\dagger$. It is easily noted that all vectors of
the form $\left|i\alpha,j\beta\right\rangle$, with $i\neq j$, are in
the null spaces of $\overline P$, $P$ and $P^\dagger$. Therefore,
$\mathcal{N}(P^{\dagger})$ is at least
$d_S^2d_E(d_E-1)$-dimensional. Where is the rest of
$\mathcal{N}(P^{\dagger})$, i.e., what is a general form of a vector
\begin{equation}
\left|c_{q,\alpha\beta}\right\rangle=\sum_{i=1}^{d_E}\xi^i_q\left|i\alpha
, i\beta\right\rangle ,\quad (\forall \mu)\;\left\langle
c_{q,\alpha\beta}| P\mu\right\rangle =0?
\end{equation}

\begin{eqnarray}
\left\langle c_{q,\alpha\beta}| P\mu\right\rangle
&=&\sum_{i,j=1}^{d_E}(\xi^i_q)^{*}\rho_E^j\left\langle i\alpha,
i\beta | j\alpha ',j\beta '\right\rangle (\mathrm{Tr}_E\mu)^{\alpha
'\beta
'},\nonumber\\
&=&(\mathrm{Tr}_E\mu)^{\alpha '\beta
'}\sum_{i=1}^{d_E}(\xi^i)^{*}\rho_E^j
\end{eqnarray}

Therefore,
\begin{equation}\label{xi}
\left\langle c_{q,\alpha\beta}| P\mu\right\rangle =0
\Longleftrightarrow \sum_{i=1}^{d_E}(\xi^i_q)^{*}\rho_E^i=0
\end{equation}
Columns $(\xi^1_q,\dots \xi^{d_{E}}_q)^\mathrm{T}$ satisfying
(\ref{xi}) constitute a $d_E-1$-dimensional space, so we conclude
that $\mathcal{N}(P^\dagger)$ is of dimension
$d_S^2d_E(d_E-1)+d_S^2(d_E-1)=d_S^2(d_E^2-1)$. Therefore, the rank
of $P$ [dimension of $\mathcal R (P)$] is $d_S^2$, so it is
isomorphic to the unit eigenspace of $\overline P$ and to the system
Liouville space. One can show that the choice
\begin{equation}\label{tilde alpha beta}
\left|\widetilde{\alpha\beta}\right\rangle
=\frac{1}{\sqrt{\mathrm{Tr}\rho_E^2}}\sum_{i=1}^{d_E}\rho_E^i\left|i\alpha
,i\beta\right\rangle
\end{equation}
indeed constitutes an orthonormal basis in $\mathcal{R}(P)$, and
that

\begin{subequations}
\begin{eqnarray}
\left\langle\widetilde{\alpha\beta}|P\mu\right\rangle=0 \quad
{\mathrm{iff}}\quad
\mathrm{Tr}_E\mu=0\\
\left\langle\widetilde{\alpha\beta}|c_{q,\alpha '\beta
'}\right\rangle=0,\quad\forall \alpha,\beta ,q,\alpha '\beta '.
\end{eqnarray}
\end{subequations}

Why was this analysis necessary? Because an uncorrelated initial
state satisfies $P\rho (0)=\rho (0)$, which means the initial
statistical operator belongs completely to $\mathcal{R}(P)$. Therefore, it
can be written in terms of the basis
$\left|\widetilde{\alpha\beta}\right\rangle$ as
\begin{equation}
\left\langle\widetilde{\alpha\beta}|\rho (0)\right\rangle
=\mathrm{Tr}_E[\rho(0)]^{\alpha\beta}\sqrt{\mathrm{Tr}\rho_E^2}=\rho_S(0)^{\alpha\beta}\sqrt{\mathrm{Tr}\rho_E^2}.
\end{equation}

\noindent In Fig. \ref{Fig 4.}, mutual relationships among the
eigenspaces of $\overline P$ and the null and image subspaces of $P$
are depicted. We obtain
\begin{widetext}
\begin{eqnarray}
\left\langle\overline{\alpha\beta}|\rho (0)\right\rangle =
\left\langle\overline{\alpha\beta}|\widetilde{\alpha\beta}\right\rangle
\left\langle\widetilde{\alpha\beta}|\rho (0)\right\rangle +
\sum_q\left\langle\overline{\alpha\beta}|c_{q,\alpha\beta}\right\rangle
\underbrace{\left\langle c_{q,\alpha\beta}|\rho
(0)\right\rangle}_{=0}
=\underbrace{\left\langle\overline{\alpha\beta}|\widetilde{\alpha\beta}\right\rangle}_{\frac{1}{\sqrt{d_E
\mathrm{Tr}\rho_E^2}}}
\underbrace{\left\langle\widetilde{\alpha\beta}|\rho
(0)\right\rangle}_{\rho_S(0)^{\alpha\beta}\sqrt{\mathrm{Tr}\rho_E^2}}.
\end{eqnarray}

The important point to note is that
$\left\langle\overline{\alpha\beta}|\rho (0)\right\rangle$ and
$\left\langle\overline{\alpha\beta}|\rho (0)\right\rangle$ are
equivalent up to the multiplicative constant
$\left\langle\overline{\alpha\beta}|\widetilde{\alpha\beta}\right\rangle=\frac{1}{\sqrt{d_E
\mathrm{Tr}\rho_E^2}}$.

We can now obtain the projection of $\rho_2(0)$ onto the
zero-eigenspace of $\overline P$ as
\begin{eqnarray}
\left\langle b_{j,\alpha\beta}|\rho (0)\right\rangle = \left\langle
b_{j,\alpha\beta}|\widetilde{\alpha\beta}\right\rangle
\left\langle\widetilde{\alpha\beta}|\rho (0)\right\rangle+
\sum_q\left\langle b_{j\alpha\beta}|c_{q,\alpha\beta}\right\rangle
\underbrace{\left\langle c_{q,\alpha\beta}|\rho
(0)\right\rangle}_{=0} =\frac{\left\langle
b_{j,\alpha\beta}|\widetilde{\alpha\beta}\right\rangle}{\left\langle\overline{\alpha\beta}|\widetilde{\alpha\beta}\right\rangle}
\left\langle\overline{\alpha\beta}|\rho (0)\right\rangle\qquad
\end{eqnarray}

\noindent Since $\left\langle
b_{i,\alpha\beta}|\widetilde{\alpha\beta}\right\rangle
=\sqrt{\frac{d_E+1-i}{(d_E-i)Tr\rho_E^2}}
\left(\rho_E^i-\frac{1}{d_E+1-i}\sum_{j=i}^{d_E}\rho_E^j\right)$,
and
$\left\langle\overline{\alpha\beta}|\widetilde{\alpha\beta}\right\rangle=\frac{1}{\sqrt{d_E
\mathrm{Tr}\rho_E^2}}$, we arrive at

\begin{eqnarray}
\left\langle b_{i,\alpha\beta}|\rho (0)\right\rangle =
\mathcal{M}^{i}\left\langle\overline{\alpha\beta}|\rho
(0)\right\rangle ,\qquad \mathcal{M}^i=
\sqrt{\frac{d_E(d_E+1-i)}{d_E-i}}\left(
\rho_E^i-\frac{1}{d_E+1-i}\sum_{j=i}^{d_E}\rho_E^j \right).\nonumber
\end{eqnarray}
\end{widetext}

Equations above fix the mapping $\rho_2(0)=\mathcal M\rho_1(0)$ from
 (\ref{rho2vsrhoS:appendix}), and explicitly embody Lindblad's
argument on the existence of a subdynamics \cite{Lindblad96}.

\section{Short-time decoherence in non-Markovian
systems}\label{sec:STE map}

In this Appendix, we formally show how to obtain the short-time
limit to the exact completely positive non-Markovian dynamical map
governing the evolution of $\rho_S$, in the form
\begin{eqnarray}\label{rho_S in the form with Gamma}
\rho_S(t)&=&\mathrm{T^c}\exp\left[\int_0^t
{\mathcal{K}}(t')dt'\right]\rho_S(0)\\
&=&
\mathrm{T^c}\exp\left\{\int_0^t dt' [
-i{\mathcal{L}}_\mathrm{{eff}}(t')-\mathcal{G}(t')]\right\}\rho_S(0),\nonumber
\end{eqnarray}
where
${\mathcal{L}}_{\mathrm{\mathrm{eff}}}(t)$ is a still
undetermined effective Liouvillian, and $\mathcal{G}(t)$ is the
dissipator term. It is well known that the form above holds for
the dynamical semigroup in the Markov approximation, where the
time-independent semigroup generator
$-i\mathcal{L}_{\mathrm{eff}}-\mathcal{G}=\mathrm{const.}$ is
of the well-known Lindblad form \cite{Lindblad76,Alicki87} that
ensures the map's complete positivity.

Here, we will perform the short-time Taylor
expansion of the exact equation (\ref{eq:dot rho_S uncorrelated}) up to
the second order in time
\begin{equation}
\rho_S(t)=\rho_S(0)+t\left(\frac{d\rho_S}{dt}\right)_0
+\frac{t^2}{2}\left(\frac{d^2\rho_S}{dt^2}\right)_0+o(t^3),
\end{equation}
and we will identify the terms in the first and second derivatives
from the desired equation (\ref{rho_S in the form with Gamma})
\begin{subequations}
\begin{eqnarray}\label{rho_S first and second derivative general}
\left(\frac{d\rho_S}{dt}\right)_0&=&\left[-i{\mathcal{L}}_\mathrm{{eff}}(0)-\mathcal{G}(0)\right]\rho_S(0),\label{dot rho_s (0)}\\
\left(\frac{d^2\rho_S}{dt^2}\right)_0&=&\left[-i\left(\frac{d\mathcal{L}_{\mathrm{eff}}}{dt}\right)_0-\left(\frac{d\mathcal{G}}{dt}\right)_0\right]\rho_S(0)\nonumber\\
&+&\left[-i{\mathcal{L}}_\mathrm{{eff}}(0)-\mathcal{G}(0)\right]^2\rho_S(0).\label{ddot
rho_s (0)}
\end{eqnarray}
\end{subequations}
with those obtained from the exact evolution described by Eq.
(\ref{eq:dot rho_S uncorrelated}).

Indeed, by using the initial conditions $R(0)=0$ and $\mathcal
V(0,0)=1$ given in Eq. (\ref{RandV}), from Eq. (\ref{eq:dot rho_S uncorrelated}) we directly obtain
\begin{eqnarray}\label{rho_S first derivative at t=0}
\left(\frac{d\rho_S}{dt}\right)_0&=&-i\left({\mathcal{L}}_{11}+{\mathcal{L}}_{12}\mathcal M\right)\rho_S(0)\nonumber\\
&=&-i\left[{\mathcal{H}}_S+\langle{{\mathcal
H}_\mathrm{{int}}}\rangle,\rho_S(0)\right].
\end{eqnarray}
Here, we have used the facts that ${\mathcal{L}}_{11}$ is generated
by the Hamiltonian ${\mathcal{H}}_S+\overline {\mathcal
H}_\mathrm{{int}}$, where $\overline {\mathcal
H}_\mathrm{{int}}=\mathrm{Tr}_E ({\mathcal H}_\mathrm{{int}})/d_E$,
while ${\mathcal{L}}_{12}\mathcal M$ is generated by the Hamiltonian
$\langle {\mathcal H}_\mathrm{{int}} \rangle - \overline {\mathcal
H}_\mathrm{{int}}$, where $\langle {\mathcal H}_\mathrm{{int}}
\rangle=\mathrm{Tr}_E \left(\rho_E {\mathcal
H}_\mathrm{{int}}\right)$ is the averaged interaction Hamiltonian
(see Appendix \ref{appendix:recipe}). Consequently,
\begin{subequations}
\begin{eqnarray}\label{Leff and dG/dt}
\mathcal{L}_{\mathrm{eff}}(0)={\mathcal{L}}_{11}+{\mathcal{L}}_{12}\mathcal
M&=&\left[{\mathcal{H}}_S+\langle{{\mathcal H}_\mathrm{{int}}}\rangle,\dots\right],\\
\mathcal{G}(0)&=&0.\hfill
\end{eqnarray}
\end{subequations}

\noindent Taking the first derivative of Eq. (\ref{eq:dot rho_S
uncorrelated}) and employing $R(0)=0$,
$\left(\frac{dR}{dt}\right)_0=i{\mathcal{L}}_{21}$, $\mathcal
V(0,0)=1$, and $\left[\frac{d\mathcal
V(t,0)}{dt}\right]_0=i{\mathcal{L}}_{22}$ [Eq. (\ref{RandV})], we
directly obtain

\begin{eqnarray}\label{rho_S second derivative at t0}
\left(\frac{d^2\rho_S}{dt^2}\right)_0=&-&\left({\mathcal{L}}_{12}{\mathcal{L}}_{21}+{\mathcal{L}}_{12}{\mathcal{L}}_{22}\mathcal
M\right)\rho_S(0)\nonumber\\
&-&{\mathcal{L}}_{11}\left({\mathcal{L}}_{11}+{\mathcal{L}}_{12}\mathcal
M\right)\rho_S(0).
\end{eqnarray}

\noindent After subtracting
$[-i\mathcal{L}_{\mathrm{eff}}(0)]^2\rho_S(0)$ from
$\left(\frac{d^2\rho_S}{dt^2}\right)_0$, what we obtain is action of
the operator
$-i\left(\frac{d{\mathcal{L}}_\mathrm{{eff}}}{dt}\right)_0-\left(\frac{d\mathcal{G}}{dt}\right)_0$
on $\rho_S(0)$. Therefore, we will introduce operator $\Lambda$ as
\begin{eqnarray}\label{Lambda definition}
2\Lambda&\equiv
&i\left(\frac{d{\mathcal{L}}_\mathrm{{eff}}}{dt}\right)_0+\left(\frac{d\mathcal{G}}{dt}\right)_0,
\end{eqnarray}
where

\begin{eqnarray}
2{\Lambda}&=&({\mathcal{L}}_{12}{\mathcal{L}}_{21}+{\mathcal{L}}_{12}{\mathcal{L}}_{22}\mathcal
M)-{\mathcal{L}}_{12}\mathcal
M\left({\mathcal{L}}_{11}+{\mathcal{L}}_{12}\mathcal
M\right).\nonumber
\end{eqnarray}
Operator ${\Lambda}$ contains essential information on the
directions of coherence loss in both non-Markovian and Markovian
systems. After a straightforward calculation, documented in Appendix
\ref{appendix:recipe}, one can obtain the matrix elements of
$\Lambda$ in the tensor-product basis of the system Liouville space

\begin{widetext}

\begin{eqnarray}\label{Lambda matrix elements}
{\Lambda}^{\alpha\beta}_{\alpha '\beta '}&=&\frac{1}{2}\sum_k
\rho_E^k
\left\{\left({\mathcal{H}}_{\mathrm{int}}^2\right)^{k\alpha}_{k\alpha
'}\delta^{\beta '}_\beta
-2\sum_{k'}\left({\mathcal{H}}_{\mathrm{int}}\right)^{k'\alpha}_{k\alpha
'}\left({\mathcal{H}}_{\mathrm{int}}\right)^{k\beta'}_{k\beta}+
\left({\mathcal{H}}_{\mathrm{int}}^2\right)^{k\beta'}_{k\beta
'}\delta^\alpha_{\alpha
'}\right\}\nonumber\\
&-&\frac{1}{2}\left\{\left(\langle
{\mathcal{H}}_{\mathrm{int}}\rangle^2\right)^{\alpha}_{\alpha
'}\delta^{\beta'}_\beta -2\langle
{\mathcal{H}}_{\mathrm{int}}\rangle^{\alpha}_{\alpha'}\langle
{\mathcal{H}}_{\mathrm{int}}\rangle^{\beta'}_\beta +\left(\langle
{\mathcal{H}}_{\mathrm{int}}\rangle^2\right)^{\beta'}_\beta\delta^{\alpha}_{\alpha
'}\right\},
\end{eqnarray}
where, for simplicity, the environmental basis is assumed to be the
eigenbasis of the environment initial statistical operator $\rho_E$. In  a
more compact form, the action of ${\Lambda}$ on any vector $\mu_S$
from the system Liouville space can be given as
\begin{eqnarray}\label{Lambda action on mu_s}
{\Lambda}\mu_S&=&\frac{1}{2}\mathrm{Tr_E}\left\{\left(\rho_E\otimes
\mu_S\right){\mathcal H}_\mathrm{{int}}{\mathcal
H}_\mathrm{{int}}-2{\mathcal H}_\mathrm{{int}}\left(\rho_E\otimes
\mu_S\right){\mathcal H}_\mathrm{{int}}+{\mathcal
H}_\mathrm{{int}}{\mathcal H}_\mathrm{{int}}\left(\rho_E\otimes
\mu_S\right)\right\}\nonumber\\
&-&\frac{1}{2}\left\{\mu_S\langle
{\mathcal{H}}_{\mathrm{int}}\rangle\langle
{\mathcal{H}}_{\mathrm{int}}\rangle -2\langle
{\mathcal{H}}_{\mathrm{int}}\rangle\mu_S\langle
{\mathcal{H}}_{\mathrm{int}}\rangle+\langle
{\mathcal{H}}_{\mathrm{int}}\rangle\langle
{\mathcal{H}}_{\mathrm{int}}\rangle\mu_S\right\}
\end{eqnarray}

An interaction Hamiltonian can always be written as
$${\mathcal{H}}_{\mathrm{int}}=\sum_i^{i_{\mathrm{max}}} f_i\otimes\varphi_i,
$$
where $f_i$ are Hermitian operators on the environment Hilbert
space, while $\varphi_i$ are Hermitian operators on the system
Hilbert space. With this form of the interaction in mind, one can
write compactly

\begin{eqnarray}\label{Lambda compact}
{\Lambda}\mu_S&=&\frac{1}{2}\sum_{i,j}\left(\langle f_i
f_j^{\dagger}\rangle - \langle f_i\rangle \langle
f_j^{\dagger}\rangle
\right)\left\{\varphi_i\varphi_j^{\dagger}\mu_S-2\varphi_j^{\dagger}\mu_S\varphi_i+\mu_S\varphi_i\varphi_j^{\dagger}\right\}\nonumber\\
&=&-\frac{1}{2}\sum_{i,j}\left(\langle f_i f_j^{\dagger}\rangle -
\langle f_i\rangle \langle f_j^{\dagger}\rangle
\right)\left\{\left[\varphi_i,\mu_S\varphi_j^{\dagger}\right]+\left[\varphi_i
\mu_S,\varphi_j^{\dagger}\right]\right\}
\end{eqnarray}
\end{widetext}

where, as before, $\langle\dots\rangle =\mathrm{Tr}_E(\rho_E\dots
)$.

>From its definition (\ref{Lambda definition}), operator $\Lambda$
satisfies
$2\Lambda=i\left(\frac{d{\mathcal{L}}_\mathrm{{eff}}}{dt}\right)_0+\left(\frac{d\mathcal{G}}{dt}\right)_0.
$ If we look at the matrix elements of the matrix [Eq. (\ref{Lambda
compact})]
$$m_{ij}=\langle f_i
f_j^{\dagger}\rangle - \langle f_i\rangle \langle
f_j^{\dagger}\rangle ,$$ we immediately note that $m_{ij}=m_{ji}^*$
due to the hermiticity of $f's$. As a result,
\begin{equation}
\frac{d{\mathcal{L}}_\mathrm{{eff}}}{dt}=0 ,\quad
\Lambda=\frac{1}{2}\left(\frac{d\mathcal{G}}{dt}\right)_0.
\end{equation}

\noindent Furthermore, $m$ is a positive-definite matrix, since for
any complex $i_{max}$ column $c=(c_1,\dots ,c_{i_{max}})^\mathrm{T}$
it holds

\begin{widetext}
\begin{eqnarray}
\langle c|m|c\rangle &=& \sum_{i,j} c^{i*} m_{ij}c^j= \sum_{i,j}
c^{i*} \left\{\langle f_i f_j^{\dagger}\rangle - \langle f_i\rangle
\langle f_j^{\dagger}\rangle \right\}c^j\\
&=&\sum_{i,j} \mathrm{Tr}_E\left\{\rho_E
\left(c^{i*}f_i\right)\left(c^{j*}f_j
\right)^\dagger\right\}-\mathrm{Tr}_E\left(\rho_E
c^{i*}f_i\right)\mathrm{Tr}_E\left(\rho_E c^{j*}f_j
\right)^*\nonumber\\
&=&\mathrm{Tr}_E\left\{\rho_E \left(\sum_{i}
c^{i*}f_i\right)\left(\sum_{i}
c^{i*}f_i\right)^\dagger\right\}-\left| \mathrm{Tr}_E\left\{\rho_E
\left(\sum_{i} c^{i*} f_i\right)\right\}\right|^2\ge 0 .\nonumber
\end{eqnarray}
\end{widetext}

\noindent The last inequality can be obtained by noting that, for
any matrix $a$,
\begin{eqnarray}
\mathrm{Tr}_E\left(\rho_E aa^\dagger\right)&=&\sum_{k,k'}\rho_E^k
\left|a^{k'}_k\right|^2\ge \sum_{k}\rho_E^k
\left|a^k_k\right|^2\nonumber\\
&\ge&\sum_{k}(\rho_E^k)^2 \left|a^k_k\right|^2\ge
\left|\sum_{k}\rho_E^k
a^k_k\right|^2\\
&=&\left|\mathrm{Tr}_E\left(\rho_E a\right)\right|^2.\nonumber
\end{eqnarray}

\noindent As a result, we conclude that \textit{$-\Lambda$ has the
form expected from the Lindblad dissipator} (it has the units of
$t^{-2}$, though, unlike the Lindblad dissipator that has the units
of $t^{-1}$).

Up to the second order in time, the generator $\mathcal K$ of the non-Markovian map (\ref{eq:integral uncorrelated}) can now be approximated as
\begin{equation}
\mathcal{K}(t)\approx-i\mathcal{L}_{\mathrm{eff}}-2\Lambda t +o(t^2),
\end{equation}
where $\mathcal{L}_{\mathrm{eff}}\equiv
\mathcal{L}_{\mathrm{eff}}(0)$ from Eq. (\ref{Leff and dG/dt}), and
$\Lambda$ is given in Eq. (\ref{Lambda matrix elements}).

\subsection{How to calculate ${\mathcal{L}}_{12}\mathcal M$ and
${\Lambda}$}\label{appendix:recipe}

\begin{widetext}
${\mathcal{L}}_{12}\mathcal M$ can be found as
\begin{eqnarray}
\left({\mathcal{L}}_{12}\mathcal
M\right)^{\overline{\alpha\beta}}_{\overline{\alpha '\beta '}}=
\sum_{j=1}^{d_E-1}\left\langle\overline{\alpha\beta} | L
|b_{j,\alpha '\beta '}\right\rangle \mathcal M^j=
\sum_{j=1}^{d_E-1}\left\langle\overline{\alpha\beta} | L
|b_{j,\alpha '\beta '}\right\rangle \frac{\left\langle
b_{j,\alpha'\beta'}|\widetilde{\alpha '\beta
'}\right\rangle}{\left\langle\overline{\alpha '\beta
'}|\widetilde{\alpha '\beta'}\right\rangle}
\end{eqnarray}

For a fixed $\alpha ,\beta$, there is a $d_E$-dimensional space
spanned by all $|i\alpha ,i\beta\rangle$. A unit operator in this
space can be written as
$\sum_{i=1}^{d_E-1}\left|b_{i,\alpha\beta}\right\rangle\left\langle
b_{i,\alpha\beta}\right|+
\left|\overline{\alpha\beta}\right\rangle\left\langle
\overline{\alpha\beta}\right|=1_{\alpha\beta}$

\begin{eqnarray}
\left({\mathcal{L}}_{12}\mathcal
M\right)^{\overline{\alpha\beta}}_{\overline{\alpha '\beta '}}&=&
\sum_{j=1}^{d_E-1}\left\langle\overline{\alpha\beta} | L
|b_{j,\alpha '\beta '}\right\rangle \mathcal M^j
=\left\langle\overline{\alpha\beta }| L \left( 1_{\alpha '\beta
'}-\left|\overline{\alpha '\beta '}\right\rangle \left\langle
\overline{\alpha'\beta'}\right|\right)|\widetilde{\alpha '\beta
'}\right\rangle
\frac{1}{\left\langle\overline{\alpha '\beta '}|\widetilde{\alpha '\beta'}\right\rangle}\nonumber\\
&=&\frac{\left\langle\overline{\alpha\beta }| L |\widetilde{\alpha
'\beta '}\right\rangle}{\left\langle\overline{\alpha '\beta
'}|\widetilde{\alpha '\beta'}\right\rangle}-
\left\langle\overline{\alpha\beta }| L |\overline{\alpha '\beta
'}\right\rangle =\sum_{i,j=1}^{d_E}\langle {i\alpha ,i\beta }| L
|{j\alpha ',j\beta '}\rangle\left(\rho_E^j-\frac{1}{d_E}\right)
\nonumber\\
&=&\sum_{i=1}^{d_E}\left( h^{i\alpha}_{i\alpha
'}\delta^{\beta}_{\beta '}-h^{i\beta
'}_{i\beta}\delta^{\alpha}_{\alpha
'}\right)\left(\rho_E^i-\frac{1}{d_E}\right)\\
&=&\sum_{i=1}^{d_E}\left[
({\mathcal{H}}_{\mathrm{int}})^{i\alpha}_{i\alpha
'}\delta^{\beta}_{\beta '}-({\mathcal{H}}_{\mathrm{int}})^{i\beta
'}_{i\beta}\delta^{\alpha}_{\alpha
'}\right]\left(\rho_E^i-\frac{1}{d_E}\right).\nonumber
\end{eqnarray}
The last line is easily obtained by showing that the contributions
from the environment Hamiltonian [$({\mathcal
H}_\mathrm{{env}})^{i\alpha}_{i\beta}=({\mathcal{H}}_E)^i_i\delta^\alpha_\beta$]
and from the system Hamiltonian [$({\mathcal
H}_\mathrm{{sys}})^{i\alpha}_{i\beta}=({\mathcal{H}}_S)^\alpha_\beta$]
vanish.

When one deals with interaction Hamiltonians of the hopping type,
i.e., those that contain an odd number of environmental
creation/annihilation operators and therefore necessarily alter the
environmental state, all
$({\mathcal{H}}_{\mathrm{int}})^{i\alpha}_{i\alpha '}=0$, and
clearly ${\mathcal{L}}_{12}\mathcal M =0$, which we used in Sec.
\ref{sec:examples}. Also, when the statistical operator is uniform
($\rho_E=\overline\rho_E$), ${\mathcal{L}}_{12}\mathcal M=0$. Note
how this term accounts for the information influx from the
environment, because it captures the deviation of the environment
statistical operator from the uniform statistical operator (the uniform statistical operator carries the maximum information entropy, i.e., environment
has no information to transmit).

In order to calculate $\Lambda$, which was defined as $2\Lambda =
{\mathcal{L}}_{12}{\mathcal{L}}_{21}+{\mathcal{L}}_{12}{\mathcal{L}}_{22}\mathcal
M-{\mathcal{L}}_{12}\mathcal
M\left({\mathcal{L}}_{11}+{\mathcal{L}}_{12}\mathcal M\right)$ in
Eq. (\ref{Lambda definition}), we should first note that
${\mathcal{L}}_{12}\mathcal
M\left({\mathcal{L}}_{11}+{\mathcal{L}}_{12}\mathcal M\right)$ is
commutator generated, i.e.,
$${\mathcal{L}}_{12}\mathcal M\left({\mathcal{L}}_{11}+{\mathcal{L}}_{12}\mathcal
M\right)=\left[\langle {\mathcal{H}}_{\mathrm{int}}\rangle-\overline
{\mathcal{H}}_{\mathrm{int}},\left[{\mathcal{H}}_S+\langle
{\mathcal{H}}_{\mathrm{int}}\rangle,\dots\right]\right]. $$ The term
${\mathcal{L}}_{12}{\mathcal{L}}_{21}+{\mathcal{L}}_{12}{\mathcal{L}}_{22}\mathcal
M$ can be rewritten as
\begin{eqnarray}
\left({\mathcal{L}}_{12}{\mathcal{L}}_{21}+{\mathcal{L}}_{12}{\mathcal{L}}_{22}\mathcal
M\right)^{\alpha\beta}_{\alpha '\beta'}&=& \langle
\overline{\alpha\beta} |
L^2-L\overline P L|\widetilde{\alpha '\beta'}\rangle \frac{1}{\langle\overline{\alpha'\beta'}|\widetilde{\alpha'\beta'}\rangle}\\
&=&\langle \overline{\alpha\beta} | L^2|\widetilde{\alpha
'\beta'}\rangle\frac{1}{\langle\overline{\alpha'\beta'}|\widetilde{\alpha'\beta'}\rangle}
-\sum_{\gamma,\sigma=1}^{d_S}\langle \overline{\alpha\beta}
|L|\overline{\gamma\sigma}\rangle\langle \overline{\gamma\sigma}
|L|\widetilde{\alpha
'\beta'}\rangle\frac{1}{\langle\overline{\alpha'\beta'}|\widetilde{\alpha'\beta'}\rangle}\nonumber\\
&=&\sum_{i,j=1}^{d_E}\langle i\alpha,i\beta | L^2|j\alpha
',j\beta'\rangle\rho_E^j\nonumber\\
&-&\frac{1}{d_E}\sum_{i,j,k=1}^{d_E}\sum_{\gamma,\sigma=1}^{d_S}\rho_E^j\langle
i\alpha,i\beta|L|k\gamma,k\sigma\rangle\langle k\gamma,k\sigma
|L|j\alpha', j\beta'\rangle , \nonumber
\end{eqnarray}
where the eigenbasis of the environment initial statistical operator
$\rho_E$ is chosen to be the environmental basis. Upon a
straightforward (and somewhat lengthy) calculation, with the only
constraint being that $[\rho_E,{\mathcal{H}}_E]=0$, which is
typically satisfied, we obtain Eq. (\ref{Lambda matrix elements}).
\end{widetext}

\section{Two Additional Examples}\label{sec:examples}
The following two examples serve to illustrate that the usefulness of the coarse-grained map (\ref{eq:Markov final})
may extend beyond the strict validity specified by (\ref{eq:validity of Markov}), and may offer a particularly simple way to identify the steady state alone from first principles.

The first example (\ref{sec:spinboson}) is analytically solvable and possesses the long-time Markovian evolution regardless of the
interaction strength. We show here that there exists a mathematical coarse-graining time $\tau$, shorter than any other timescale in the system or environment, so that
the exact long-time Markovian evolution coincides with that obtained from the short-time evolution by coarsening over $\tau$ (\ref{eq:Markov final}).

On the second example (\ref{sec:Jaynes-Cummings}), we show that relaxation towards the correct equilibrium state is easily obtained by using  (\ref{eq:Markov final}) (or equivalently by employing Theorem 2 in
Sec. \ref{sec:Decoherence in the presence of a "memoryless" environment}).

\subsection{Spin-boson model with pure
dephasing}\label{sec:spinboson}
One of the few analytically solvable
\cite{Lidar01,Palma96,Duan98,Grifoni97,Thorwart04,Emary04} open
system problems is that of a two-level system coupled to a
dephasing-only boson bath, with the relevant Hamiltonians are given
by
\begin{eqnarray}
{\mathcal{H}}_S&=&\frac{\omega}{2}\sigma_z,\;
{\mathcal{H}}_E=\sum_{\vec{q}}\Omega_q\left(b^\dagger_{\vec q}
b_{\vec q}+\frac{1}{2}\right),\nonumber\\
{\mathcal{H}}_{\mathrm{int}}&=&\sum_{\vec q} \sigma_z\left\{
g(\Omega_{\vec q})b_{\vec q}+g(\Omega_{\vec q})^{*}b_{\vec
q}^\dagger\right\}.
\end{eqnarray}
Here, $\sigma_z$ is the Pauli matrix, $b_{\vec q}^\dagger$ and
$b_{\vec q}$ and the boson creation and annihilation operators of
the $q$-th boson mode, respectively, $\pm\omega/2$ are the system
energy levels (divided by $\hbar$), and $\Omega_{q}$ is the boson
mode frequency. The boson modes are initially in a thermal state
with $\langle n_q\rangle=\langle b^\dagger_q b_q\rangle
=\frac{1}{\exp(\hbar\Omega_q /k_B T)-1}$. Because of the interaction
linear in environment creation/annihilation operators, $\langle
\mathcal H_{\mathrm{int}}\rangle=0$, so
${\mathcal{L}}_{S}={\mathcal{L}}_{\mathrm{eff}}$:
\begin{eqnarray}
&{\mathcal{L}}_{\mathrm{eff}}={\mathcal{L}}_S=\omega\left[\begin{array}{cccc}
0 & 0 & 0 & 0 \\
0 & 1 & 0 & 0 \\
0 & 0 & -1 & 0 \\
0 & 0 & 0 & 0 \end{array}\right],\end{eqnarray} \noindent where the
rows/columns are ordered as $1=\left|+\right\rangle\left<+\right|,
2=\left|+\right\rangle\left<-\right|,
3=\left|-\right\rangle\left<+\right|,
4=\left|-\right\rangle\left<-\right|$ ($\pm$ refer to the
positive/negative (upper/lower) energy state). Operator $\Lambda$
can be calculated according to (\ref{Lambda matrix elements}) as
\begin{eqnarray}\label{lambda spin-boson}
{\Lambda}&=&{\lambda_d}\left[\begin{array}{cccc}
0 & 0 & 0 & 0 \\
0 & 1 & 0 & 0 \\
0 & 0 & 1 & 0 \\
0 & 0 & 0 & 0 \end{array} \right],\nonumber\\
{\lambda_d}&=&2\sum_{\vec q}|g(\Omega_{\vec q})
|^2\coth{\left(\frac{\hbar\Omega_q}{2k_BT}\right)}\nonumber\\
&=&2\int_0^{\infty} d\Omega
\mathcal{D}(\Omega)|g(\Omega)|^2\coth{\left(\frac{\hbar\Omega}{2k_BT}\right)},
\end{eqnarray}
where $\mathcal{D}(\Omega)$ is the density of boson states.

${\mathcal{L}}_{S}$ and ${\Lambda}$ obviously commute, and their
common zero eigenspace [$\mathcal N(\Lambda)=\mathcal
N({\mathcal{L}}_{\mathrm{eff}})$] contains all density matrices with
zero off-diagonal elements. This means that, for a given initial
statistical operator, the off-diagonal matrix elements will decay to zero
while the diagonal elements remain unchanged:
\begin{eqnarray}\label{spin boson Markov}
\left[\rho_S(t)\right]_{++}&=&\left[\rho_S(0)\right]_{++},\;
\left[\rho_S(t)\right]_{--}=\left[\rho_S(0)\right]_{--},\nonumber\\
\left[\rho_S(t)\right]_{+-}&=&\left[\rho_S(0)\right]_{+-}e^{-i\omega
t-\lambda_d\tau t},\\
\left[\rho_S(t)\right]_{-+}&=&\left[\rho_S(0)\right]_{-+}e^{+i\omega
t-\lambda_d\tau t}, \nonumber
\end{eqnarray}
The steady state will be determined by simply annulling the
off-diagonal elements. This is the correct steady state, as shown in
the exact solution \cite{Breuer02}.

Instead of $\exp(-\lambda_d\tau t)$, in the exact solution
decoherence is seen through the term $\exp[-\Gamma(t)]$, where the
$\Gamma(t)$, the decoherence exponent, behaves as
\begin{equation}
\Gamma(t)=\int_0^\infty d\Omega
2\mathcal{D}(\Omega)|g(\Omega)|^2\coth{\left(\frac{\hbar\Omega}{2k_BT}\right)}\frac{\sin^2(\Omega
t/2)}{(\Omega/2)^2}.
\end{equation}

\noindent For short-times, $\Gamma(t)\approx \lambda_d t^2$, as
should be expected, because we know our expansion (\ref{eq:short time expansion of K}) is
exact up to the second order in time. In the long-time limit for
$\Gamma(t)$, only the low frequency contributions survive, since
$\lim_{t\rightarrow\infty}\frac{\sin^2(\Omega t/2)}{(\Omega/2)^2
t}=\pi\delta(\Omega)$, so
\begin{equation}
\Gamma(t\rightarrow\infty)=t\lim_{\Omega\rightarrow 0} 2\pi
\mathcal{D}(\Omega)|g(\Omega)|^2\coth{\left(\frac{\hbar\Omega}{2k_BT}\right)}.
\end{equation}
We need to match this long-time behavior of $\Gamma(t)$ with our
coarse-grained term $\lambda_d\tau t$, in order to obtain $\tau$.
\begin{equation}\label{tau spin boson}
\tau=\lim_{t\rightarrow\infty}\frac{\Gamma (t)}{\lambda_d
t}=\frac{\lim_{\Omega\rightarrow 0} 2\pi
\mathcal{D}(\Omega)|g(\Omega)|^2\coth{\left(\frac{\hbar\Omega}{2k_BT}\right)}}{2\int_0^{\infty}
d\Omega
\mathcal{D}(\Omega)|g(\Omega)|^2\coth{\left(\frac{\hbar\Omega}{2k_BT}\right)}}.
\end{equation}

\noindent Let us consider the example of an Ohmic bath (e.g., page
228 of Ref. \onlinecite{Breuer02}), with
$D(\Omega)|g(\Omega)|^2=\frac{1}{4}\Omega\exp(-\Omega/\Omega_c)$ and
$\Omega_c$ being a density-of-states cutoff frequency. Typically,
$\hbar\Omega_c\gg k_BT$. In the numerator, one can approximate
$\coth{\left(\frac{\hbar\Omega}{2k_BT}\right)}\approx
\frac{2k_BT}{\hbar\Omega}$, while the $\coth{}$ function in the
denominator is always greater than 1, yielding
\begin{equation}
\tau
<\left(\frac{k_BT}{\hbar\Omega_C}\right)\frac{2\pi}{\Omega_c}\ll\frac{2\pi}{\Omega_c}.
\end{equation}
Being typically the largest frequency scale in the full $SE$
problem, $\Omega_c$ sets the shortest physical timescale. Clearly,
$\tau$ is even shorter than the period associated with $\Omega_c$,
which justifies our use of the short-time expansion and subsequent
coarse-graining.

Note the long-time behavior $\exp(-t/\tau_T)$ of the decoherence
term $\Gamma$, where $\tau_T=\hbar/\pi k_B T$ is the thermal
correlation time. However, our time $\tau$ is the
\textit{mathematical} coarse-graining time, which is very short. The
relationship between the correct physical correlation loss time and
the mathematically appropriate time is
\begin{equation}
\tau=(\lambda_d\tau_T)^{-1}.
\end{equation}

\subsection{Jaynes-Cummings model in the rotating wave
approximation}\label{sec:Jaynes-Cummings}

The Jaynes-Cummings Hamiltonian in the rotating-wave approximation
\cite{Jaynes63,Meystre88,Hussin05,VanWonderen95} describes the decay
of a two-level system in the presence of a single boson mode of
resonant frequency. The relevant Hamiltonians are

\begin{eqnarray}\label{Jaynes Cummings Hamiltonian}
{\mathcal{H}}_S&=&\frac{1}{2}\omega\sigma_z ,\;
{\mathcal{H}}_E=\omega \left(b^\dagger
b+\frac{1}{2}\right),\nonumber\\
{\mathcal{H}}_\mathrm{int}&=&g\left(b^\dagger \sigma_{-}+b
\sigma_{+}\right).
\end{eqnarray}

\noindent Here, $\sigma_z,$ $\sigma_{+}=\left(\sigma_x
+i\sigma_y\right)/e$, and $\sigma_{-}=\left(\sigma_x
-i\sigma_y\right)/2$ are the Pauli matrices, $b^\dagger$ and $b$ are
the boson creation and annihilation operators, respectively,
$\pm\omega/2$ are the system energy levels (in units of frequency)
and $\omega$ is also the boson mode frequency, and $g$ is a
parameter measuring the interaction strength. The boson mode is
initially in a thermal state with $\langle n\rangle=\langle
b^\dagger b\rangle =\frac{1}{\exp(\hbar\omega /k_B T)-1}$. As in the
spin-boson example, ${\mathcal{L}}_{\mathrm{eff}}={\mathcal{L}}_S$
because of the interaction linear in environment
creation/annihilation operators:
\begin{eqnarray}
{\mathcal{L}}_{\mathrm{eff}}={\mathcal{L}}_S=\omega\left[\begin{array}{cccc}
0 & 0 & 0 & 0 \\
0 & 1 & 0 & 0 \\
0 & 0 & -1 & 0 \\
0 & 0 & 0 & 0 \end{array}\right].
\end{eqnarray}

Operator $\Lambda$ can be calculated according to Equation  (\ref{Lambda matrix elements}) as
\begin{eqnarray}
{\Lambda}=\frac{g^2}{2}\left[\begin{array}{cccc}
2\langle n\rangle +2 & 0 & 0 & -2\langle n\rangle \\
0 & 2\langle n\rangle +1 & 0 & 0 \\
0 & 0 & 2\langle n\rangle +1 & 0 \\
-2\langle n\rangle -2 & 0 & 0 & 2\langle n\rangle \end{array}
\right].
\end{eqnarray}

\noindent ${\mathcal{L}}_S$ and ${\Lambda}$ commute, and we
immediately note two common one-dimensional eigenspaces: $\rho_{+-}$
is associated with the ${\mathcal{L}}_S$ and ${\Lambda}$ eigenvalues
$\omega$ and $g^2(2\langle n\rangle +1)/2$, respectively, while
$\rho_{-+}$ is associated with the eigenvalues $-\omega$ and
$g^2(2\langle n\rangle +1)/2$.

On the other hand, the space spanned by $|+\rangle\langle +|$ and
$|-\rangle\langle -|$ is the null space of
${\mathcal{L}}_{\mathrm{eff}}$. Solving the eigenproblem of
${\Lambda}$ reduced to this space gives

\begin{eqnarray}
\det\left[\begin{array}{cc}
g^2\left(\langle n\rangle +1\right)-\lambda & -g^2\langle n\rangle \\
-g^2\left(\langle n\rangle +1\right) & g^2\langle n\rangle-\lambda
\end{array}\right]=0,\\
\lambda =0\quad \mathrm{and}\quad \lambda=\lambda_d\equiv
g^2\left(2\langle n\rangle +1\right).\nonumber
\end{eqnarray}
An eigenvector $\mu^{0}=(\mu^{0}_{++},\mu^{0}_{--})^\mathrm{T}$
corresponding to the zero eigenvalue of the matrix $\Lambda$ is
characterized by
\begin{equation}\label{mu^0}
\mu^{0}_{--}=\mu^{0}_{++}\frac{\langle n\rangle +1}{\langle
n\rangle}.
\end{equation}
If we are looking for a statistical operator that belongs to the zero
eigenspace of $\Lambda$, it also has to satisfy the constraint of
the unit trace, which fixes
\begin{equation}\label{mu^0 w trace=1}
\mu^{0}_{++}=\frac{\langle n\rangle}{2\langle n\rangle +1},\quad
\mu^{0}_{--}= \frac{\langle n\rangle +1}{2\langle n\rangle +1}.
\end{equation}
One recognizes these components as the \textit{thermal equilibrium
values} of the population of the upper and lower level of our
two-level system, respectively (see, for instance, p. 149 of Ref.
\onlinecite{Breuer02}). Therefore, by seeking the steady state in
$\mathcal N(\Lambda)\cap\mathcal N({\mathcal{L}}_{\mathrm{eff}})$,
we have obtained the physically correct result.



\end{document}